\documentclass[useAMS,usenatbib]{mn2e}
\usepackage{graphicx}

\title[Supernova 1999dn]
{The Type Ib SN~1999dn: One Year of Photometric and Spectroscopic Monitoring.
\thanks{Based on observations collected at the European Organisation for Astronomical Research in the Southern Hemisphere, Chile (ESO programmes 64.H-0604 and 65.H-0292),
at the Italian 3.58m Telescopio Nazionale Galileo and the William Herschel Telescope (La Palma, Spain), and at the Copernico telescope (Asiago, Italy).}}

\author[S. Benetti et al.]{S.~Benetti,$^{1}$\thanks{E-mail: stefano.benetti@oapd.inaf.it}  
M.~Turatto,$^2$  S.~Valenti,$^3$ A.~Pastorello,$^3$ E.~Cappellaro,$^1$ M.T.~Botticella,$^{3}$ 
\newauthor {F.~Bufano,$^{1}$ F.~Ghinassi,$^{4}$ A.~Harutyunyan,$^{4}$ C.~Inserra,$^2$  A.~Magazz\`u,$^{4}$ F.~Patat,$^5$}
\newauthor {M.L.~Pumo,$^{2}$ S.Taubenberger$^{6}$}  \\
$^1$Osservatorio Astronomico di Padova, vicolo dell'Osservatorio 5, I-35122, Padova, Italia \\
$^2$Osservatorio Astrofisico di Catania, Via S. Sofia 78, I-95123, Catania, Italia \\
$^3$Astrophysics Research Centre, School of Mathematics and Physics, Queen's University Belfast, BT7 1NN, UK \\
$^4$Telescopio Nazionale Galileo, Fundaci\'on Galileo Galilei - INAF, Rambla Jos\'e Ana Fern\'andez P\'erez, 7, 38712 Bre\~na Baja, TF - Spain \\
$^5$European Southern Observatory, Karl Schwarzschild-Str. 2, 85748, Garching bei M\"unchen, Germany \\
$^6$Max-Planck-Institut f\"ur Astrophysik, Karl Schwarzschild Str. 1, 85741 Garching bei M\"unchen, Germany
}

\def\ni{$^{56}$Ni}
\def\co{$^{56}$Co}
\def\fe{$^{56}$Fe}

\def\J{SN~1993J}

\def\dn{SN~1999dn}

\def\ax{SN~2008ax}
\def\d{SN~2008D}
\def\ex{SN~1999ex}
\def\y{SN~2007Y}
\def\kms{km\,s$^{-1}$}
\def\mcento{mag\,(100d)$^{-1}$}
\def\Ha{H$\alpha$}
\def\Hb{H$\beta$}

\def\Mv{M$_{\rm V}$}

\def\M{M$_{\odot}$}

\def\ebv{E(B--V)}
\def\tbb{T$_{\rm bb}$}
\def\vph{v$_{\rm ph}$}
\def\lam{$\lambda$}

\begin{document}
\date{Received ................; accepted ................}


\maketitle

\begin{abstract}
Extensive optical and near-infrared (NIR) observations of the type Ib supernova 1999dn are presented, covering the first year after explosion. These new data turn this object, already considered a prototypical SNIb, into one of the best observed objects of its class. The light curve of  \dn\/ is mostly similar in shape to that of other SNeIb but with a moderately faint peak (M$_V=-17.2$ mag). From the bolometric light curve and ejecta expansion velocities, we estimate that about $0.11$\M~ of \ni~ were produced during the explosion and that the total ejecta mass was $4-6$\M~ with a kinetic energy of at least $5\times 10^{51}$ erg. The spectra of \dn\/ at various epochs are similar to those of other Stripped Envelope (SE) SNe showing clear presence of H at early epochs. The high explosion energy and ejected mass, along with the small
flux ratio [CaII]/[OI] measured in the nebular spectrum, together with the lack of signatures of dust formation
and the moderate metallicity environment is not inconsistent with a single massive progenitor (M$_{\rm ZAMS}\geq23-25$ \M) for SN~1999dn. 


\end{abstract}


\section{Introduction}
In the Supernova (SN) taxonomy, type Ib Supernovae (SNIb) are defined as the subclass 
of Core-Collapse explosions (CC-SNe) whose early-time spectra are
characterized by strong HeI lines \citep[e.g.][]{wheeler85,turatto07}.
CC-SNe are thought to descend from massive progenitors
(M~$> 7-8$ \M) and include also the most frequent,  H-dominated
type II SNe (SNII) as well as type Ic SNe (SNIc), which appear deprived of both H and He.
The physical connection among these subtypes is provided 
by their location almost exclusively in spiral galaxies \citep[e.g.][]{hakobyan08}
and, in particular, by their association to massive star formation regions \citep[e.g.][]{vandyk99,anderson}.
An additional evidence of a link between the different subtypes of  CCSNe came with the discovery of objects, called type IIb,
that metamorphose from type II at early stages to type Ib later on.
The prototype of this subclass is \J\/ in M81, 
one of the best studied SNe at all wavelengths ever \citep[e.g.][]{barbon95,richmond96b}.

While early time spectra of CC-SNe can be very different as a consequence of the different configurations
of the progenitors at the moment of their explosions, late time spectra of all
CC-SNe are consistently similar with strong emission lines of neutral and singly ionized 
O and Ca, in addition to H Balmer lines for SNII.
The standard scenario is that SNIb descend from massive stars that have lost their H envelope through strong winds or mass transfer to a companion
\citep{heger03}. If in addition to the H
also the He envelope has been removed, then the SN will appear of type Ic.
For this reason SNIb, Ic and IIb are often referred to as Stripped Envelope (SE) SNe.

With the improved quality of data, the differences between type Ib and Ic have shown to be  subtle,
and classifications often controversial. For instance, in SN~1994I, early on considered as the prototypical
SNIc,   has been found possible signature of He \citep{filippenko95,clocchiatti96};
the type Ib \ex\/  was characterized by weak optical HeI but strong HeI 
\lam10830, \lam20581 lines in the near-IR \citep{hamuy02};
the peculiar  SN~2005bf  \citep{folatelli06} and SN 2008D \citep{mazzali08,modjaz09} 
underwent a metamorphosis from a broad line type Ic at early 
times to a typical type Ib at later epochs. The sharp distinction between 
the two classes seems therefore to be replaced by a continuity of properties in He 
abundances and/or excitation.

The study of stripped-envelope SNIb/c has received fresh impetus in the past decade
because of the association
of some of them, in particular the most energetic SNIc (hypernovae),
with GRBs of long duration \citep{galama98, hjorth03,malesani04}.
More recently a few SNIc have been associated with less energetic X-ray flashes
\citep[][]{fynbo04,modjaz06,pian06}. An X-ray flash was also detected in the type Ib \d,
which was attributed either to shock break-out at the star
surface \citep{soderberg08} or to the effect of mildly 
relativistic jets due to the collapse of a 30 \M\/ star to a black hole \citep{mazzali08}.

Despite this renewed interest the objects with detailed observations remain few. 
In particular, it is not well ascertained if  SNIb do exist as a distinct class or if
there is a uniform distribution of objects from SNII to SNIc with decreasing H (and He) content in the outer layers.
In this context, \dn\/ is interesting because it has been adopted several times in the past to describe the average properties of SNIb \citep[e.g.][]{branch02, chornock10}.

\begin{figure}
\includegraphics[height=80 mm]{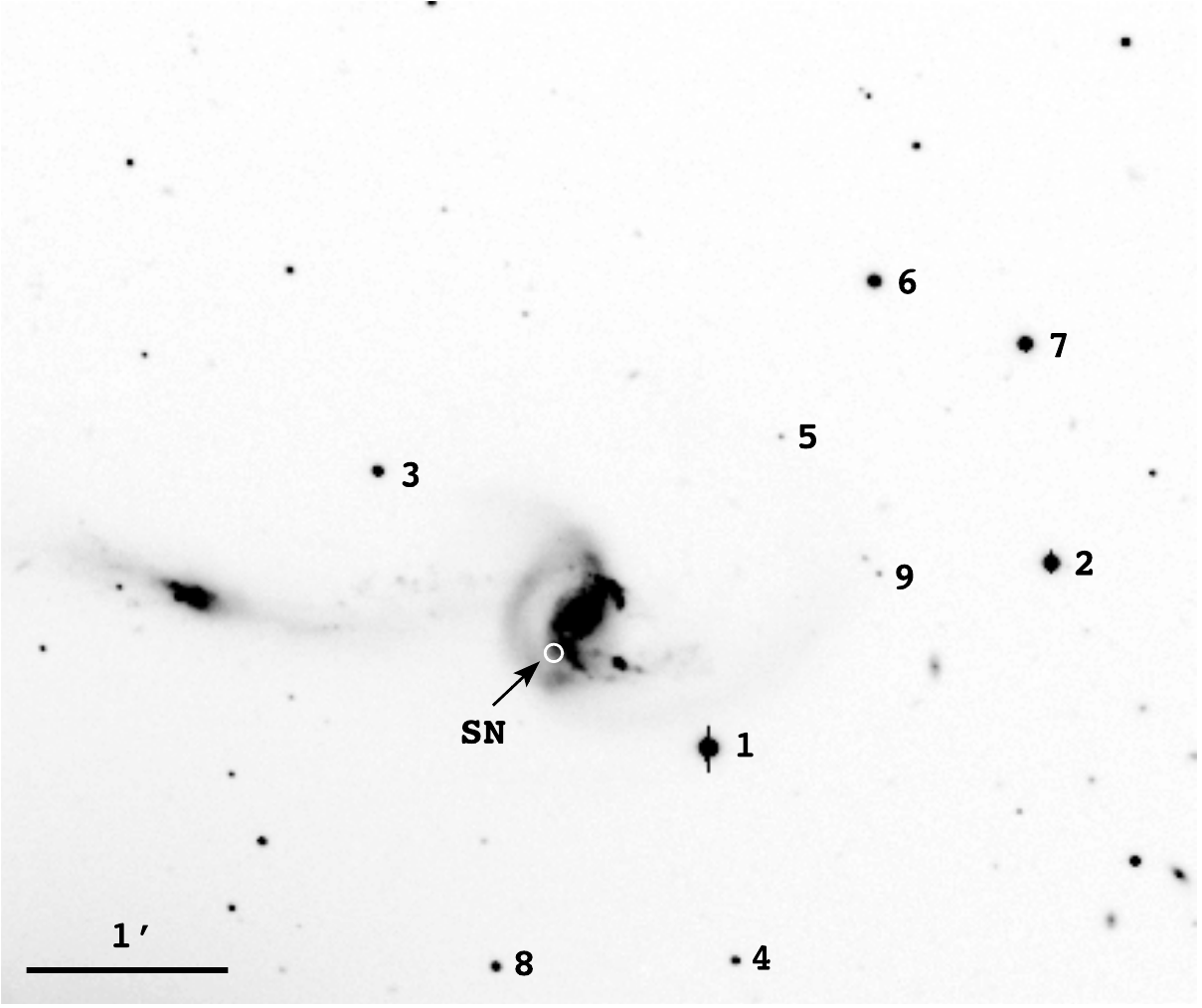}
\caption{SN~1999dn in NGC~7714 and the sequence of local reference stars (cfr. Tab.~1). The image is a
V frame taken with TNG+Dolores on June 28, 2000. North is up, east to the left.} 
\label{fig:sn99dn2}
\end{figure}

\dn\/ was discovered by \citet{qiu} on two
unfiltered CCD images obtained on Aug. 19.76 and 19.82 UT, respectively,
in the Wolf-Rayet galaxy NGC 7714. The SN position is
R.A.(2000.0)=23$^h$36$^m$14$^s$.7; Dec(2000.0)= +02$^o$09$'$
08$\arcsec$.8,  9.9$\arcsec$E, 9.4$\arcsec$S of the galaxy nucleus \citep{qiu},
in a region of steep background variation (Fig. \ref{fig:sn99dn2}). 
The parent galaxy, NGC~7714, is classified as  SBb peculiar, and identified by \citet{weed} as a prototypical starburst galaxy. 
A second SN (2007fo), has been recently discovered 
2.5$\arcsec$W, 12.4$\arcsec$N of the galaxy nucleus  \citep{KhandrikaLi07}, which showed similarly prominent He~I lines and was also classified
as SNIb \citep{Desroches07}. Another highly reddened SN candidate was
detected  2$\arcsec$W, 5$\arcsec$N of the galaxy nucleus on UKIRT archival
K-band images taken on 1998, Dec. 5 
\citep[but not in H band, i.e. (H--K)$>1.2;$][]{mattila}.

A series of spectra of \dn\/ has been taken soon after the discovery 
by different groups which classified the SN as type Ia \citep{qiu2}
and as type Ic because of the lack or weakness of He lines \citep{ayani99, turatto99}. 
Few days later the HeI lines emerged
and the SN was re-classified as a SNIb/c by \citet{pasto99}. 
A week later the HeI lines strengthened again
making the spectra of this SN very similar to those of other type Ib's.

Due to the early discovery, \dn\/ was observed by several groups
\citep{deng, matheson}. It soon became
one of the best--observed SNIb and a test case for extensive modeling \citep{deng,branch02,ketchum08,james10}.
In this paper we present original data collected at La Silla, La Palma and Asiago,
and analyze them together with published material.
The joint set of observations gives  good multicolor photometry and dense spectral sampling, 
starting one week before maximum up to over two months.
The SN has been recovered at late time in imaging  \citep{vandyk03} and
spectroscopy  \citep{tauben09}.

\section{Observations} \label{obs}

UBVRIJHK$'$ photometry of \dn\/ was obtained at ESO-La Silla and
TNG-La Palma. Ten photometric nights were used to calibrate a local sequence of stars
through comparison with photometric standard stars \citep{land}. In turn, the local
sequence was used to calibrate observations obtained during
non-photometric nights. The magnitudes of the local sequence, labeled
in Fig. \ref{fig:sn99dn2}, are reported in Tab. \ref{seq}. 
The estimated errors (mean standard deviation) are reported in parentheses.
Due to the small field of view of Arnica only star 1 was always included in the NIR frames, which then is the reference to check the calibration of the JHK$'$ photometry.
The averages of the measurements of this stars in three nights are reported in the footnote of Tab. \ref{seq}
along with their standard deviations. The moderate dispersions of the measurements and their consistency with  the  2MASS Point Source Catalogue (difference $<0.2 mag$)
\citep{2mass}   add confidence to the photometric calibration.

\begin{table}
\caption{Magnitudes of the stars of the local sequence identified in
Fig.~1. The photometric errors are in parentheses (in units of $10^{-2}$ mag).} \label{seq}
\begin{tabular}{lccccc}
\hline
star& U & B & V & R & I \\
\hline   
1$^*$        &15.20(05)&14.73(03)&13.84(01)&13.43(03)&12.86(02)\\
2        &15.76(06)&15.45(04)&14.65(02)&14.20(02)&13.80(03)\\
3        &17.91(13)&17.35(06)&16.38(02)&15.83(01)&15.36(04)\\
4        &19.36(16)&18.58(08)&17.62(03)&17.07(09)&16.56(06)\\
5        &22.82:   &21.57(09)&19.95(07)&18.73(03)&17.44(04)\\
6        &16.19(02)&16.20(04)&15.44(05)&14.97(03)&14.55(06)\\
7        &15.68(05)&15.69(03)&15.03(02)&14.64(02)&14.27(02)\\
8        &19.83(09)&18.43(09)&17.02(04)&16.13(04)&15.47(11)\\
9        &22.71(09)&21.31(09)&19.72(04)&18.70(05)&17.58(12)\\
\hline
\end{tabular}
\begin{flushleft}
$^*$ J$=12.25\pm0.05$,  H$=11.50\pm0.10$, K$'=11.53\pm0.10$
\end{flushleft}

\end{table}

The new photometric measurements of the SN are listed in
Tab.~\ref{obs_tab}. Observations were obtained on 19 different epochs
up to one year after explosion. Data reduction followed
standard procedures making use of a PSF fitting technique for the SN
measurements.  The mean photometric
errors, estimated with artificial stars experiments, are given in
parentheses.

\begin{table*}
\caption{Photometric measurements for SN~1999dn. The photometric errors are in parentheses (in units of $10^{-2}$ mag).}\label{obs_tab}
\begin{tabular}{lcccccccccl}
\hline
date&JD$^*$& U & B & V & R & I & J & H & K$'$ & instr.\\
\hline
25/08/99&51415.71 & 16.93(06)& 16.80(05)& 16.18(04)& 15.96(04)& 15.82(06)&         &         &         & DF \\
28/08/99&51418.79 & 16.93(06)& 16.81(05)& 16.10(04)& 15.80(04)& 15.72(06)&         &         &         & DF \\
03/09/99&51424.77 & 17.23(06)& 17.01(05)& 16.24(04)& 15.92(04)& 15.65(06)&         &         &         & DF \\
06/09/99&51427.77 & 17.62(06)& 17.15(05)& 16.30(05)& 15.95(06)& 15.66(06)&         &         &         & DF \\
11/09/99&51433.49 & 18.59(06)& 17.71(05)& 16.59(04)& 16.11(05)& 15.69(05)&         &         &         & OIG \\
14/09/99&51435.76 & 18.71(06)& 17.80(03)& 16.64(03)& 16.15(03)& 15.81(04)&         &         &         & OIG \\
15/09/99&51436.50 & 18.97(09)& 17.87(05)& 16.76(03)& 16.28(03)& 15.85(03)&         &         &         & EF2 \\
18/09/99&51439.77 & 19.28(09)& 18.14(05)& 16.91(03)& 16.41(03)& 15.95(03)&         &         &         & EF2 \\
20/09/99&51441.58 &          &          &          &          &          &15.38(15)&15.16(10)&14.92(30)& ARN \\
03/10/99&51455.48 &          &          &          &          &          &15.78(10)&15.50(10)&15.31(20)& ARN \\
07/10/99&51459.47 & 19.99(09)& 19.01(06)& 17.81(05)& 17.09(05)& 16.63(05)&         &         &         & OIG \\
02/11/99&51485.35 & 20.14(15)& 19.59(06)& 18.32(06)& 17.56(06)& 16.88(06)&         &         &         & OIG \\
03/11/99&51485.58 &          & 19.53(05)& 18.32(03)& 17.57(05)&          &         &         &         & DF \\
04/11/99&51486.52 &          & 19.47(06)& 18.30(03)& 17.58(04)& 16.95(05)&         &         &         & DF \\
10/12/99&51523.33 & 20.28(10)& 19.89(08)& 18.82(07)& 18.02(07)& 17.30(07)&         &         &         & OIG \\
27/12/99&51540.32 &          & 19.94(08)& 19.11(07)& 18.20(07)& 17.59(07)&         &         &         & OIG \\
27/12/99&51540.37 &          &          &          &          &          &17.93(20)&17.46(25)&17.64(30)& ARN \\
28/06/00&51723.69 &          &          & 22.57(35)& 20.78(20)&          &         &         &         & Dol \\
31/08/00&51788.60 &          &          &$\ge 23.15(70)$&21.93(50)&          &         &         &         & EF2 \\
\hline
\end{tabular}
\begin{flushleft}
$^*$ 2400000+ \\
DF = ESO/Danish 1.5m telescope + DFOSC\\
OIG = Telescopio Nazionale Galileo + OIG CCD Camera\\
EF2 = ESO 3.6m telescope + EFOSC2\\
ARN = Telescopio Nazionale Galileo + ARNICA IR Camera\\
Dol = Telescopio Nazionale Galileo + Dolores\\
\end{flushleft}
\end{table*}

The journal of the spectroscopic observations
(Tab. \ref{spec_tab}) gives for each spectrum the date of observation (col.1), the
phase relative to the adopted maximum (cfr. Sect. \ref{sect:phot}; col.2), 
the instrument (col.3), the exposure time (col.4), the
wavelength range (col.5) and the resolution derived from the average
FWHM of the night-sky lines (col.6). In order to improve the signal-to-noise ratio, the
average of the two spectra taken on 3 and 4 September 1999 is shown
in Fig.~\ref{spec_fig}, after
checking that there was negligible evolution.
In some spectra the telluric absorptions have not been removed because of the lack
of suitable standard stars with sufficient signal--to-noise ratio.
In other cases, it has been impossible to remove the contamination by the underlying HII region
whose lines can be either under--  (e.g. day +370) or over--subtracted (day +0.8).
The absolute flux calibration was verified against
the B, V and R photometry. In case of disagreement, the spectra were
corrected by applying a constant factor. In fact the slit was normally aligned along the parallactic angle, in order to minimize atmospheric differential light losses \citep{fil82}.

\section{Photometry} \label{sect:phot}

\begin{figure*}
\includegraphics[width=17cm, angle=0]{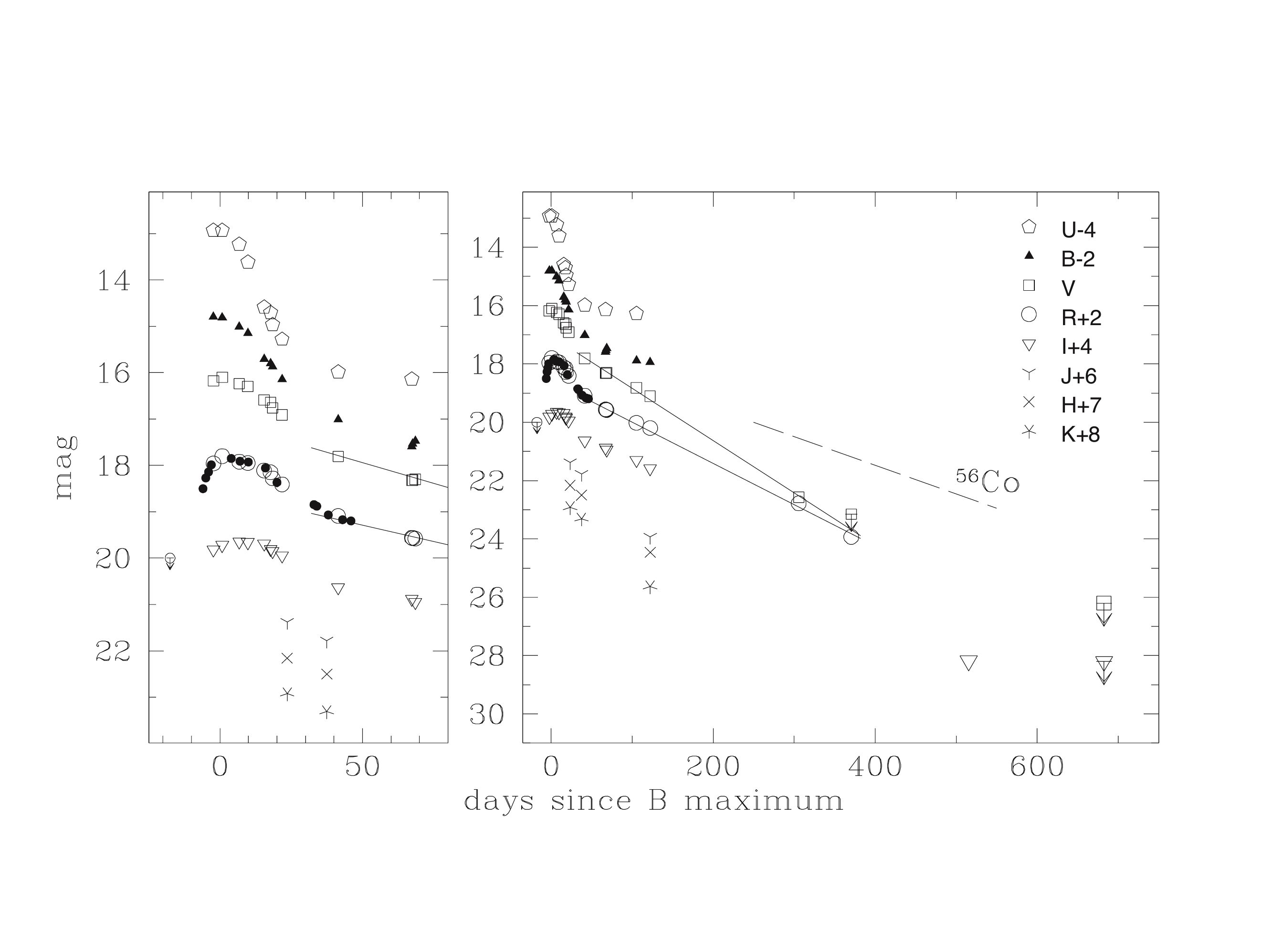}
\caption{ UBVRIJHK$'$ light curves of SN 1999dn. The left panel is a zoom 
around the epoch of maximum, while the whole evolution is shown in the right panel. 
The R band observations by \citet{matheson} are reported as small filled circles, 
shifted to match our R magnitudes.  The prediscovery unfiltered limit by \citet{qiu} 
is reported in the R band scale (open circle + arrow). Late time ($> 500$d) F550W and 
F814W observations by \citet{vandyk03} are reported in the V and I scale, respectively, as large symbols.
To guide the eye, the late time V and R observations have been fitted by straight lines,
while a dashed line is drawn to show the slope of \co\/ decay.}
\label{fig:99dn_lc_lo}
\end{figure*}

The UBVRIJHK$'$ light curves of SN~1999dn are shown in
Fig.~\ref{fig:99dn_lc_lo}.
The figure includes also the R-band data from Fig.~27 of
\citet{ matheson} (shifted by 15.85 mag to match our observations), the late time observations
of \citet{vandyk03} and the pre-discovery limit by \citet{qiu}.
Only in the R band the light curve has been well monitored before and around
maximum. A low order polynomial fit of all points around the peak provides 
JD$^{\rm R}_{max} = 2451420.5 \pm 0.5$ (30 Aug. 1999) at R$_{\rm max}=15.85\pm0.05$ mag,
in good agreement with the estimate by \cite{matheson} (31 Aug. 1999).
In the other bands the observations started a few days later and
the uncertainties on the epochs of maxima are larger (cfr. Tab.~\ref{maindata}).
Following common convention we adopt the epoch 
of B maximum  (JD 2451418.0) as reference which is 3.5 days before
the reference epoch used by \citet{matheson} and, later, by \citet{branch02}, \citet{ketchum08} and \citet{james10}.

The peak of the light curve is asymmetric with a steep rise to maximum followed 
by a  slow decline for about 10 days and a faster decrease afterwards.
The contrast between maximum and inflection point
as well as the width of the maximum light seems to
be color dependent, with shorter wavelengths having narrower peaks and larger
magnitude differences. A color dependence is visible also in the 
decline rate during the late radioactive tail,
longer wavelengths having steeper slopes (cfr. Tab.~\ref{maindata}).
Only three epochs of NIR photometry are available:  two during the early post peak decline 
and one in the radioactive tail. Though scanty the NIR photometry has been very
useful for the construction of the bolometric flux (Sect.\ref{sec:bol}).

\begin{table}
\caption{Spectroscopic observations of SN 1999dn} \label{spec_tab}
\begin{tabular}{cccccc}
\hline
\hline
     Date & phase$^*$ & inst.$^{**}$ & exp. & range & resol. \\ &
     (days)& & (min) & (\AA) & (\AA) \\
\hline
25/08/99 & --2.3  &  DF   &  45   & 3600-9000 &     9      \\
28/08/99 & +0.8  &  DF   &  45   & 3500-9000 &     9      \\
03/09/99 & +6.8  &  DF   &  45   & 3500-9000 &     9      \\
09/09/99 & +12.5 &  WHT  &  45   & 3200-7550 &     3.5    \\
14/09/99 & +17.5 &  EK    &  30   & 3470-7470 &     18     \\
14/09/99 & +17.8 &  EF2  &  15   & 3400-7450 &     14     \\
3/11/99  & +67.6   &  DF   & 120   & 3500-9000 &     10     \\
4/11/99  & +68.6 &  DF   & 120   & 3500-9000 &     10     \\
31/08/00 & +370.7  &  EF2  &  60   & 6000-10250&     12     \\
\hline
\end{tabular}

$^*$relative to the estimated epoch of B maximum (JD=2451418).

$^**$See note to Table~2 for coding, plus WHT = WHT 4.2m telescope + ISIS and EK = Asiago 1.8m telescope + AFOSC.
\end{table}

\begin{figure}
\includegraphics[width=9.5cm]{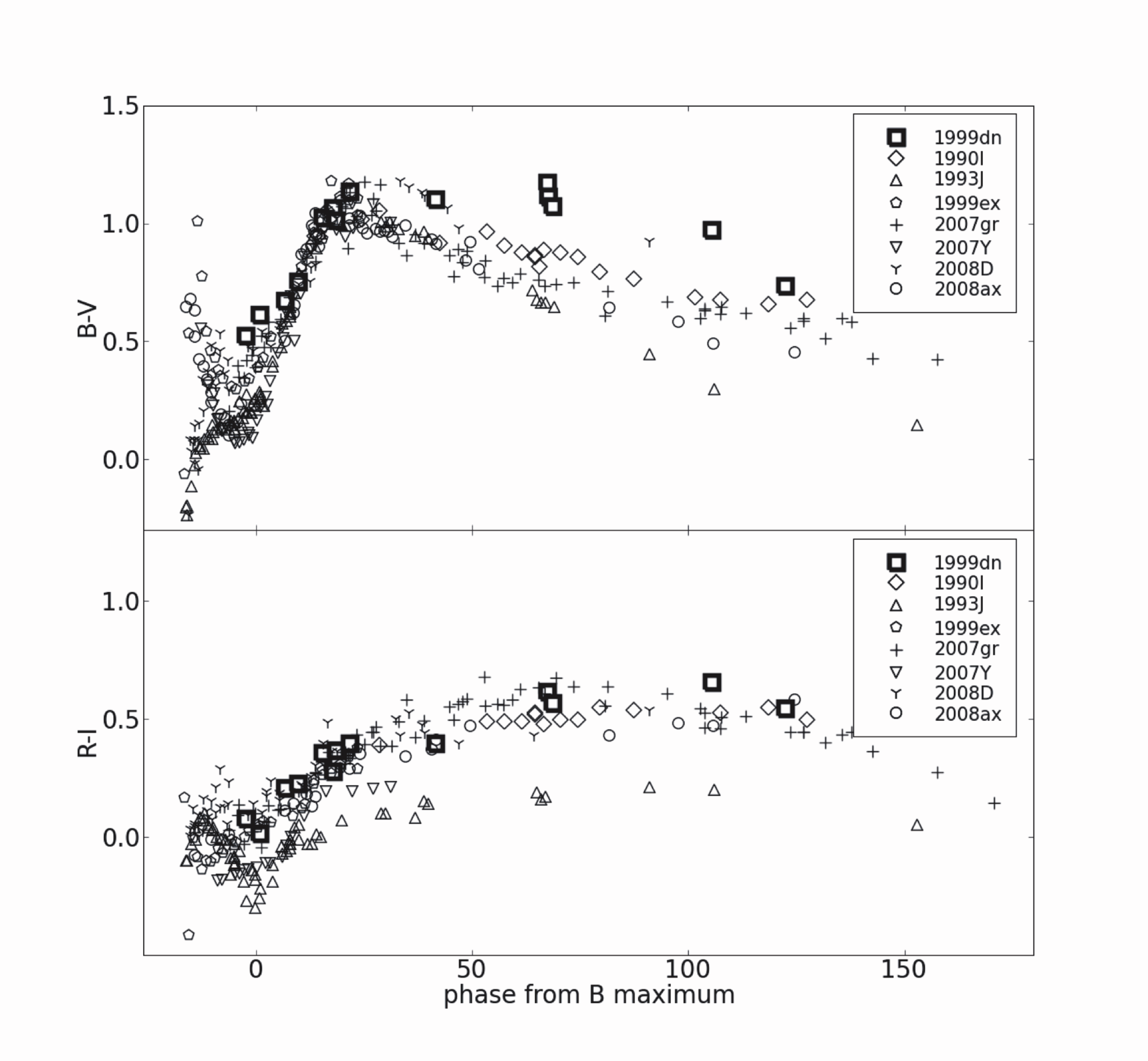}
\caption{Top panel: Comparison of the B--V color curve of \dn\/ with those
of other SE--SNe. The curves have been dereddened with the values reported in Tab.~\ref{comp_sample}
and a standard extinction law.
Bottom panel: as above, but for the observed R--I color curve (r--i for \y). }
\label{fig:col-conf}
\end{figure}

A comparison of the colour evolutions of \dn\/ with several other SNIb/c is illustrated in Fig.~\ref{fig:col-conf}.
All objects have been dereddened according to values reported in Tab~\ref{comp_sample}.
An overall similarity is found among all objects of this variegated class but in the pre-maximum phase.
At maximum light the B--V colors are between +0.1 to +0.6 mag and become redder afterward, reaching a maximum 
value (0.9 to 1.2 mag) at about 20d because of the cooling due to expansion.
Thereafter the B--V colors slowly turn bluer due to the progressive increase of the
emission line strengths.
A good coverage before maximum is available only for the helium rich SNe 1993J, 1999ex, 2008D and 2008ax,
which show very different behaviors.
Whereas the blue-band light curves of the type IIb \J\/ showed an early peak attributed to the 
emergence of the shock breakout,
\ax, the other SNIIb of the sample, showed evidence of its emergence only in the very early UV light curve \citep{rom09}, but not in the optical
\citep{pasto08,tauben10} and thus the color curve had an opposite behavior, starting red and turning bluer.
Already one week before maximum the B--V curves of both objects
reach similar values (B--V$=+0.15$ mag) and evolve similarly afterwards. 
The pre-maximum difference may be attributed to the fact that
\ax\/ had a more compact progenitor and a lower density wind than \J\/ \citep{chevalier10}.
An early peak in the blue light curves is also visible in SN 2008D and possibly in \ex.
Indeed both objects show early B--V colors comparable to \J. 
However, the blue color peak is very short in \ex\/ and the evolution extremely fast, with the B--V color
reddening by over 1 mag in only 3 days and resembling the general trend described for \ax\/ thereafter.  
The behavior of \ex\/ leaves room for the presence of early, short shock breakout signatures also in other objects.

The colors of \dn\/ are available only from around maximum, but the spectral evolution in the week before (cfr. Sect.~\ref{sec:spec}) indicates that the SED has become bluer and bluer from day --6 to maximin light, which suggests a color evolution for SN 1999dn more similar to SNe 1999ex and 2008ax than to SN 1993J. After reaching a maximum value of B--V similar to all other SNIb/c, \dn\/ remained unusually red up to over 100 days. The adoption of an higher extinction correction might reduce the difference (cfr. Sect. \ref{sect:red}).

The R--I color evolutions of SE--SNe are smoother and remain confined between $-0.3$ mag at maximum and $+0.7$ mag later on. All objects show similar evolution and \dn\/ is in the troop.
The outlier here is \J, which remains bluer at all epochs.

\begin{table}
\scriptsize
\caption{ Main data  of SN~1999dn\label{maindata}}
\begin{tabular}{lcc}
\hline
position (2000.0)$^a$	&23$^h$36$^m$14.81$^s$	&+02$^o$09$^m$08$^s$.4 \\
parent galaxy			& \multicolumn{2}{c}{NGC~7714,  SBb pec$^b$ with starburst$^c$ }\\
offset wrt  nucleus 		& 9.9$\arcsec$E		& 9.4$\arcsec$S 	\\
adopted distance modulus	&  $\mu=32.95\pm0.11$	&	\\
SN heliocentric velocity	&  $2630\pm150$ \kms	&	\\
adopted reddening		& E$_{MW}$(B--V$)=0.052^d$	& E$_{tot}$(B--V$)=0.10\pm0.05$\\
\end{tabular}
\begin{tabular}{lccc}
\hline
	& peak time		& peak observed	& peak absolute  \\
	&  (JD--2451000)	&  magnitude		&  magnitude	\\
U	& $418\pm2$		&$16.9\pm0.1$		&$-16.6\pm 0.5$	\\
B	& $418\pm2$		&$16.8\pm0.1$		&$-16.6\pm 0.4$\\
V	& $419\pm1$		&$16.1\pm0.1$		&$-17.2\pm 0.4$\\
R	& $420.5 \pm 0.5^e$&$15.85\pm0.05$&$-17.35\pm 0.40$ \\
I	& $424\pm1$		&$15.60\pm0.05$	&$-17.55\pm 0.35$\\
uvoir	& $419.5\pm0.5$	& 		&L$_{bol}=2.0 \times 10^{42}$ erg s$^{-1}$  \\
	&				&				&				\\
rise to max & $\sim12$ days	& 			&			\\
explosion day & $\sim406$ & $\sim16$ Aug. 1999 &			\\
\end{tabular}
\begin{tabular}{lcccc}
\hline
	&late time decline	& interval	&  late time decline	& interval \\
	&  \mcento		& days	& \mcento		  	& days     \\
U	&	0.37			& 67 --105&				&	\\
B	& 	0.81			& 67--122 &				&	\\
V	& 	1.79			& 67--306 &	1.43			& 67--122\\
R	& 	1.44			& 67--370 &	1.17			& 67--122$^f$\\
I	& 	1.64			& 67--515 & 	1.19			& 67--122\\
uvoir&       1.49			& 67--122 &      				&	\\
\end{tabular}
\begin{tabular}{lc}
\hline
M(\ni)			&  0.11 \M \\
M(ejecta)			& 4--6 \M  \\
explosion energy	&  5.0--7.5 $\times 10^{51}$  erg\\
\hline
\end{tabular}
\begin{flushleft}
a- \citet{qiu}, b- http://leda.univ-lyon1.fr, c- \citet{weed}, d- \citet{sch}, e- 30 Aug. 1999, f- our data only
\end{flushleft}
\end{table}

\subsection{Reddening and absolute magnitudes} \label{sect:red}

The interstellar NaID lines arising both  
from our and the parent galaxy are well seen in the low-resolution spectra
at about  5893 and 5945 \AA, respectively. 
Therefore the reddening of SN~1999dn cannot  be neglected.
\citet{sch} give a galactic reddening E$_{MW}(\rm B-V)= 0.052$ mag
in the direction to NGC~7714. We have measured the EWs of the NaID components
in the spectra of SN~1999dn on the spectrum of highest signal--to--noise and resolution
(WHT on 9 Sept., res. 3.5 \AA) 
and found an EW$_{\rm MW}$(NaID)$\sim0.51$ \AA\ and
EW$_{\rm N7714}$(NaID)$\sim0.45$ \AA.
These values are in good agreement with the average of EW measurements
performed on all other available spectra with no evidence of 
significant time evolution.
Assuming that the gas/dust properties in NGC~7714 and the Galaxy are the same, 
we estimate a total redening E$_{tot}(\rm B-V)= 0.098$ mag.
A similar value E$_{tot}(\rm B-V)= 0.131$ mag is obtained by using the average relation 
between E(B--V) and EW${\rm (NaID)}$ \citep{turatto03}.  
Finally, the match of the color curves of \dn\/ to those other SE SNe (cfr. Sect.~\ref{sect:phot})
suggests a consistent, though formally slightly higher reddening, \ebv$\sim0.15$.
Throughout this paper we adopt for \dn\/ a total reddening E$_{tot}(\rm B-V)= 0.10\pm0.05$ mag
(A$_V = 0.31$ mag). 

Since there is no direct measurements of the distance to NGC~7714, we use the Hubble's law. 
From the wavelength of the interstellar
NaID absorption features we derive a recession velocity of $2630 \pm
150$ \kms, which is consistent with the heliocentric radial
velocity reported by LEDA\footnote{http://leda.univ-lyon1.fr} ($2797\pm16$ \kms). 
LEDA provides also a velocity corrected for the Local Group infall into the Virgo cluster $v_{Vir}=2798$ \kms.
Adopting H$_0=72$ \kms Mpc$^{-1}$, we derive a distance modulus $\mu=32.95$ mag which
is used throughout this paper\footnote{A similar value ($\mu=32.84$) is obtained adopting the distance 
relative to the Virgo cluster in the 220 model \citep{kraan}  with D(Virgo)=15.3 Mpc \citep{freedman}.}.
%

Adopting the above mentioned values for extinction and distance, we
obtain the following absolute magnitudes at maximum $M_{\rm U} = -16.6\pm 0.5$,  $M_{\rm B} = -16.6\pm 0.4$, 
$M_{\rm V} = -17.2\pm 0.4$, $M_{\rm R} = -17.35\pm 0.40$ and $M_{\rm I} = -17.55\pm 0.35$,
where in the computation of the errors we adopted an uncertainty  of $\pm250$ \kms\/ on the Hubble distance modulus due to possible peculiar motion.

We note that, after rescaling to a common distance scale, our determination of 
$M_{\rm V}$ is 0.5 mag brighter than that derived by \citet{richardson06}, but in good agreement with the mean value of $M_{\rm R} = -17.01\pm 0.41$ given by \citet{li10} for a sample of six SNIb, SN 1999dn included. In comparison to the subsamples studied in Li et al., \dn\/ remains about 0.80
mag brighter than their unweighted mean for the $normal$ SE SNe (SNIc+Ib+IIb).

\begin{figure}
\includegraphics[width=9cm]{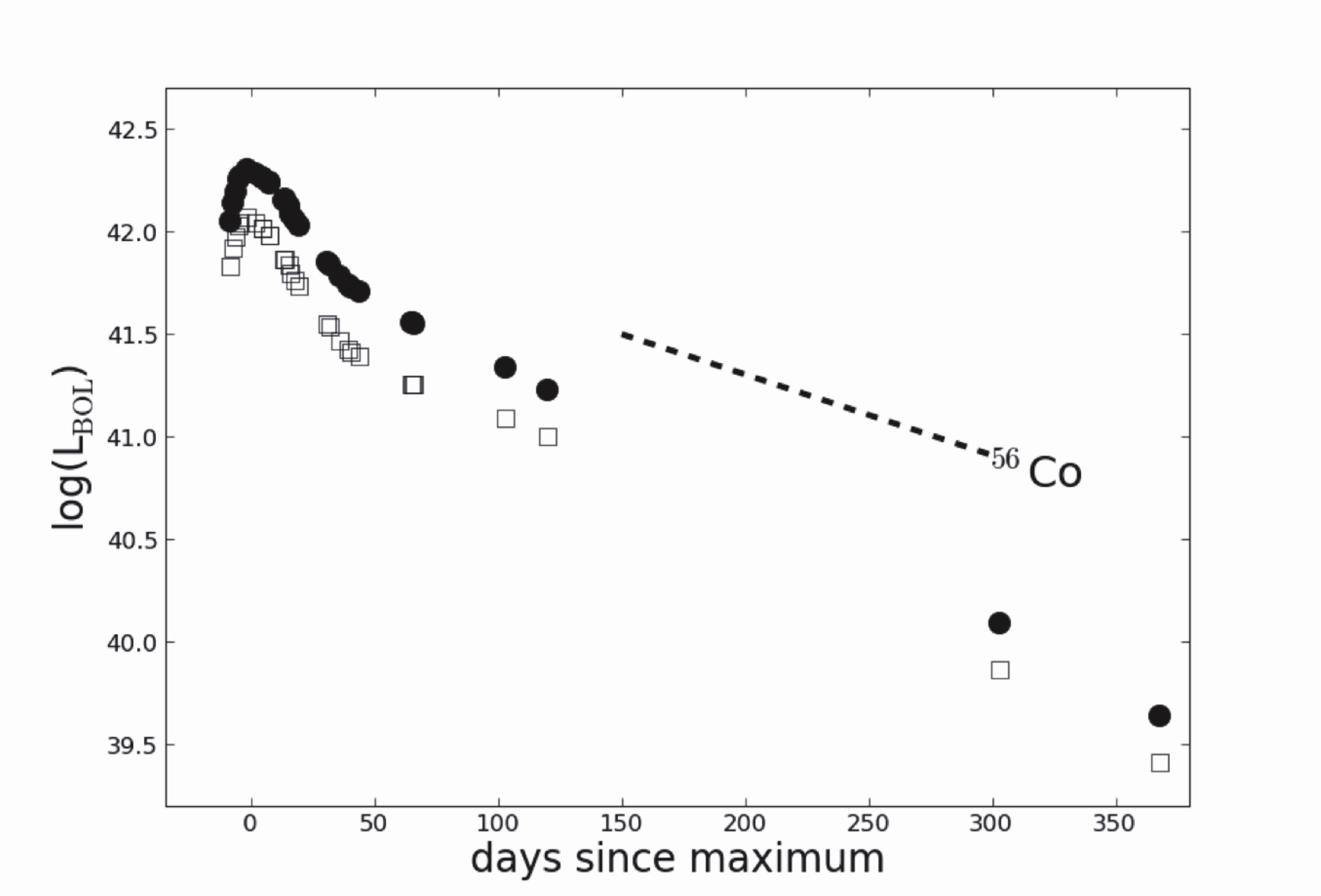}
\caption{UBVRIJHK$'$ bolometric (filled circles) and UBVRI (open squares) light curves of SN 1999dn.
The points on the rising branch and those after 4 months
past maximum are largely based on extrapolations (assuming constant colors) and should be regarded as uncertain.
The slope of \co\/ to \fe\/ decay is also displayed for comparison. }
\label{fig:99dn-bolo}
\end{figure}

\begin{table*}
\caption{SE SNe used for color and bolometric luminosity comparison.}\label{comp_sample}
\begin{tabular}{lccccccccl}
\hline
SN	&	Host Galaxy	& SN type	& E$_{tot}$(B--V) & (m-M) & main references\\  
\hline
1999dn    & NGC 7714     & Ib            & 0.10       &  32.95    & Sect.~\ref{sect:red} of this paper \\
		&			&		&		&		&	\\
1990I	& NGC 4650A	& Ib		& 0.16	& 32.90	& \cite{elmhamdi04} \\                  
1993J 	& NGC 3031	& IIb		& 0.30	&  27.80	& \cite{barbon95,richmond96b}\\   
1994I	& NGC 5194	& Ic		& 0.45 	&  29.75 	& \cite{richmond96a,dessart08} \\    
1999ex	& IC 5179	         & Ib		& 0.28 	& 33.19	&  \cite{stritzinger02}\\                 
2004aw	& NGC 3997	& Ic		& 0.37	& 34.17	& \cite{tauben06}\\                       
2005bf	& MCG +00-27-5	& Ib/c	&  0.05 	& 34.46 & \cite{folatelli06}\\                    
2007Y 	& NGC 1187	& Ib		& 0.11	&  31.13	& LEDA, \cite{stritzinger09}\\                   
2007gr 	& NGC 1058 	& Ic		& 0.09	&  29.84 	& \cite{valenti08a,hunter09} \\              
2008D	& NGC 2770	& Ib		& 0.66	& 32.29	& \cite{mazzali08,modjaz09}\\            
2008ax	& NGC 4490	& IIb	& 0.40	& 29.92	& \cite{pasto08,tauben10}\\            
\hline
\end{tabular}
\end{table*}

\subsection{Bolometric light curve} \label{sec:bol}

With the available photometry it has been possible to build the UBVRIJHK$'$ bolometric light curve up to 1yr (Fig.~\ref{fig:99dn-bolo}).
When magnitudes in a bandpass were not available in a given night, the values were linearly interpolated. 
The light curves with a short temporal extension, e.g. those in the NIR, were extrapolated
assuming  constant colors.
For this reason the bolometric flux during the rising branch and those at the latest epochs are based mainly on R band observations and should be regarded as uncertain. 
The photometry was corrected for extinction using R$_V=3.1$.
At the effective wavelengths of each filter monochromatic fluxes were computed;
these were then integrated from the U to K$'$ bands using the trapezoid approximation, and converted to luminosity.
The peak of the bolometric light curve is reached between the V and the R maxima on JD$^{\rm bol}_{max} = 2451419.5 \pm 0.5$ at a luminosity L$_{bol}=2.0 \times 10^{42}$ erg s$^{-1}$.

The contribution of the NIR bands is substantial at all the epochs in which
NIR data are available (L$_{JHK}$/L$_{uvoir}=$ 0.5, 0.5 and 0.4 at day 20, 40 and 121, respectively).
This is more than a factor two larger than the value derived for SNIb 2007Y \citep{stritzinger09} and for SN~2008D \citep{modjaz09}, but similar to values derived for SNIIb  2008ax \citep{tauben10}.
In the photospheric phase, the optical+NIR SED deduced from SN~1999dn photometry is consistent  with a blackbody energy distribution with temperatures as derived from optical spectra (cfr. Sect. \ref{sec:spec}).
The decline rate between 67d and 122d is 1.49 \mcento, close to the
average value of type Ia SNe.
This value is significantly larger than the decline rate of 0.98 \mcento\/ predicted if all the 
energy from the decay of  \co\/ into \fe\/  was fully thermalized.
No significant slope variation is observed up to day 370 (${\gamma}_{\rm uvoir}=1.65$  [67d-305d]).

\begin{figure}
\includegraphics[width=9.5cm]{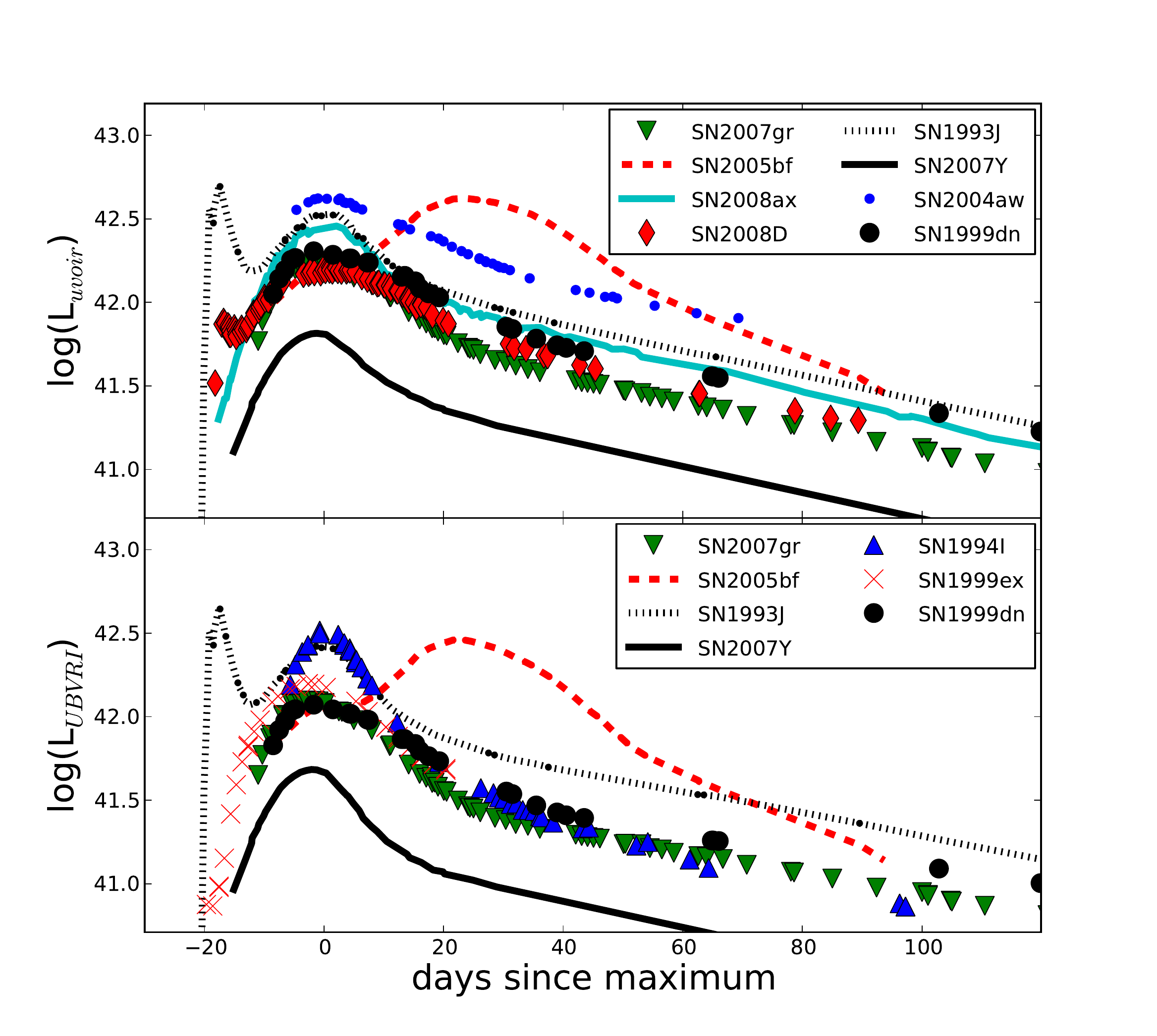}
\caption{Comparison of the bolometric light curve of \dn\/ with those of
other type Ib/c SNe. The phase is from B maximum. The comparison based on the whole optical-NIR (uvoir) domain is in the
top panel, that based on the UBVRI bands (uBgVri for SN 2005bf, UuBgVri for \y) in the bottom panel.
The SN distances and reddening are reported in Tab.~\ref{comp_sample}. 
Note that with the adopted distance and reddening SN~1994I is among the brightest objects.
Small misalignments in the epochs of maxima are due to different choices of the reference epochs.}
\label{fig:bolo-conf}
\end{figure}

In Fig.~\ref{fig:bolo-conf} we compare the bolometric light curve of \dn\/ with those of other SE SNe.
Unfortunately not all have coverage from optical to NIR 
wavelengths. The comparison has been done both for the extended $uvoir$ (UBVRIJHK$'$, top panel)
and the optical-only  (UBVRI, bottom panel) bolometric curves.\\
The rise to maximum can be very different. 
The type IIb SN~1993J shows an early bright spike about 20d before
a broader secondary maximum. For the other SNIb caught early-on, SN~1999ex (bottom panel), the bolometric light curve  shows just an hint  of the shock breakout. \\
For the peculiar SN~2005bf a slow rise to a first maximum
is observed followed about 25d later by a second brighter maximum.
The type Ib \y\/ and the type Ic SNe 1994I, 2004aw and  2007gr were not detected sufficiently early after the explosion and do not show any feature in the rising branch.
The first reliable photometry of \dn\/ has been obtained  6 days before B maximum (8.5d before R maximum), while the available pre-discovery limit does not put stringent constraints on the date of explosion. The direct comparison with other SNIb/c suggests a rise time (relative to B maximum) similar to SN~2007gr, i.e. $11.5\pm2.0$  days \citep{hunter09}.

The contribution of the JHK$'$ bands to the bolometric flux of \dn\/ stands out: while in the UBVRI domain it  has almost the same luminosity as the type Ic SN~2007gr, in the  $uvoir$ it outpasses it by about 0.12 dex.\\
The asymmetric peak of SN 1999dn, with a relatively fast rise and slow decline, reminds of the behavior of SN 2007gr, but with even slower decline. After maximum the bolometric light curve remains relatively broad and closely resembles in shape that of SN~2004aw, but  remains about 0.3 dex fainter at all epochs.
This simple comparison indicates that adopting the Arnett model \citep{arnett82} at the early times, when the diffusion approximation is valid, the amount of \ni\/ synthesized in the explosion of \dn\/ is between 0.10 and 0.15 \M\/ (M$_{Ni}$(2007gr)=0.076 \M, \citet{hunter09}; M$_{Ni}$(2004aw)=0.30 \M, \citet{tauben06}). 
Broad light curves are indicative of large diffusion time and hence of 
either a large ejected mass or a small kinetic energy. The comparison of the expansion velocity of SN 1999dn (cfr. Sec. \ref{sec:Ha}) with other SE SNe suggests that the long diffusion time is due to a
large ejecta mass of \dn\/, much larger than that of the narrow-peak SN~1994I (cfr. Fig.~\ref{fig:bolo-conf}).
A more detailed discussion of the explosion parameters is given in Sect.~\ref{sec:disc}.

\begin{figure*}
\includegraphics[width=18cm]{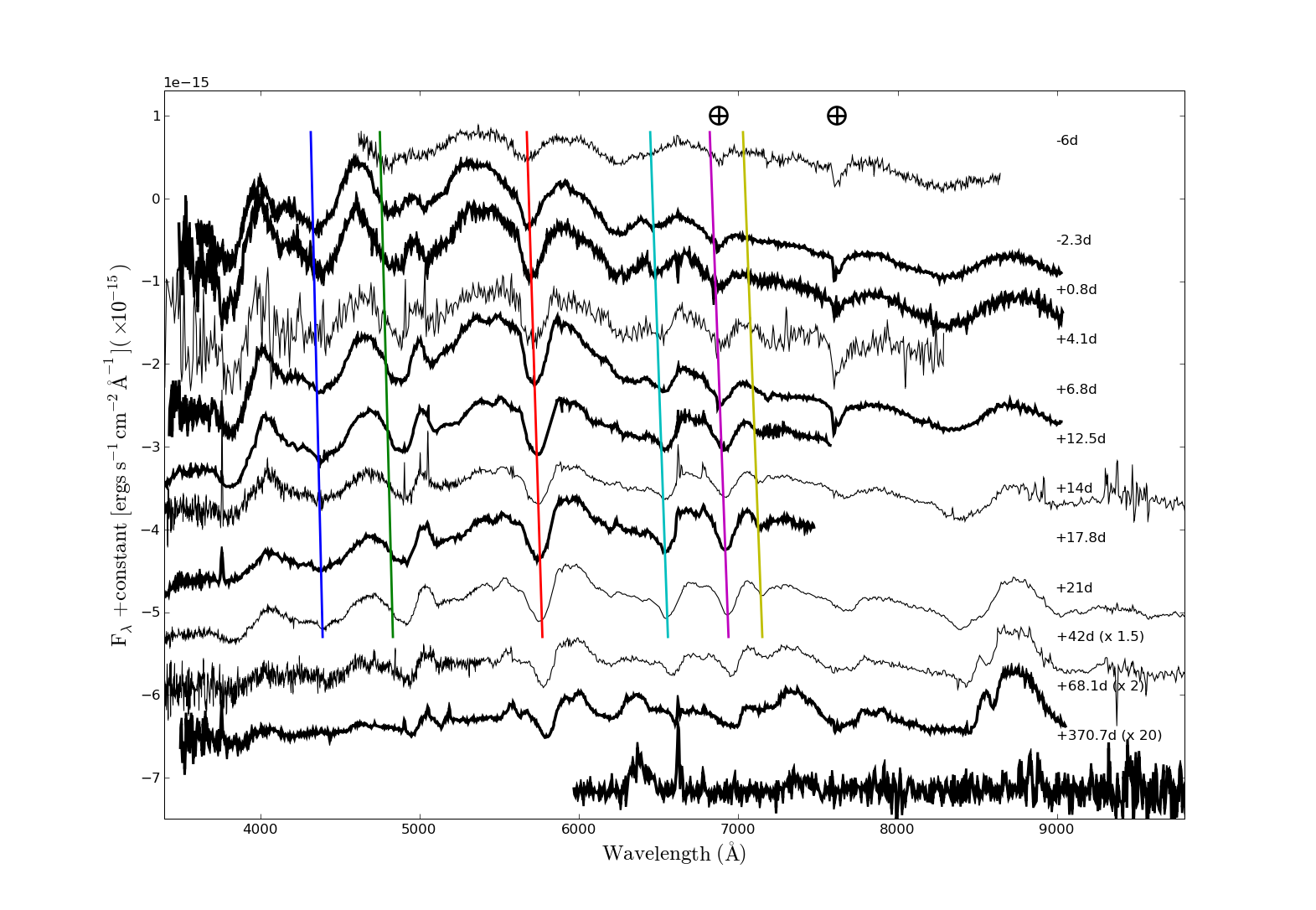}
\caption{The overall spectral evolution of SN~1999dn
(including spectra by \citealp{deng} and \citealp{matheson}  as thin lines).  The wavelength is in the observer rest frame and the flux is not corrected for reddening. The ordinate refers to the top spectrum; other spectra are shifted downwards with respect to the previous by $1.2\times 10^{-15}$ (second spectrum) and $0.6\times 10^{-15}$ erg s$^{-1}$ cm$^{-2}$ \AA$^{-1}$ (others). The spectra taken on Nov.~3 and 4 have been averaged, while for Sept.~14 only the EFOSC2 spectrum is displayed.
Vertical lines correspond to the positions of HeI $\lambda4472$, $\lambda4921$, $\lambda5876$, $\lambda6678$, $\lambda7065$ and $\lambda7281$, with expansion velocities of 13000 \kms\/ at the earliest epoch  and then slowing down with time.  
Each spectrum is labelled with the phase with respect to the B maximum (JD=2451418.0).
For clarity the last three spectra have been multiplied by the factors reported in parentheses. Telluric features are marked with the Earth symbol.}
\label{spec_fig}
\end{figure*}

\section{Spectroscopy} \label{sec:spec}

The entire spectroscopic evolution of \dn\/ from about 1 week before maximum to over 1yr is illustrated in Fig.~\ref{spec_fig}, thanks to the spectra listed in Tab. \ref{spec_tab} and those published  by \citet{deng} and \citet{matheson}.
The SN evolution is therefore well sampled at all crucial phases. We stress again that the phases adopted here, are relative to the date of B maximum (JD 2451418.0) and differ from those used in previous papers (cfr. Sect.~\ref{sect:phot}).

The {\bf first spectrum (--6d, Aug.21)} has been subject to deep scrutiny.
\citet{deng} have modeled it with SYNOW using a black-body temperature \tbb=7500 K and photospheric velocity \vph=16000 \kms. 
Besides the broad absorptions at 4750 and 4950 \AA\/ (restframe) due to FeII, and at  about 8200 \AA\/ due to CaII IR, the spectrum is dominated by the HeI $\lambda5876$ and $\lambda7065$ which contribute to the strong 5620 \AA\/ and 6820 \AA\/ features.
The broad absorption at about 6200 \AA\/ has attracted the attention of Deng et al. and other modelers and will be extensively discussed in Sect.~\ref{sec:Ha}.
This spectrum has been modeled also by \citet{branch02} who used SYNOW with line optical depths varying as $v^{-n}$ and $n=8$ instead of $e^{-v/v_e}$ ($v_e=1000$ \kms) as in \citet{deng}. The spectrum was well reproduced with \tbb$=6500$ K and \vph$=14000 $\kms and the same contributing ions.
\citet{ketchum08} modeled this and the other spectra around maximum with their non-LTE code PHOENIX trying both a standard solar and a three times higher metallicity.
Although  they adopted a smaller extinction (\ebv=0.052 mag) and later explosion epoch  (12 days before Aug.31) with respect to the values that we now consider more realistic (cfr. Tab.~\ref{maindata}), their fit  is good also for the non-thermal HeI lines. The other features were identified with CaII, FeII and OI.
Similar results were obtained also by \citet{james10}, who correct the spectrum only for galactic extinction (Baron, private communication).
Their model parameters (T$_{mod}$=6000 K and v$_0=11000$\kms) are slightly smaller than the values 
(\tbb$=6700\pm200$ K and \vph$=12800$\kms) we derive by fitting the de-reddened spectrum. Note that our \tbb~ is a color temperature, different from T$_{mod}$  which is much closer to an effective temperature, while the afore mentioned SYNOW \tbb~ has little physical significance \citep{deng}.

Our new observations provide a good coverage of the epoch of maximum with spectra on {\bf days --2.3 and +0.8}. At this phase the object has become bluer and hotter, reaching \tbb$=9100$ and 7800 K, respectively.
The line contrast is now higher with more pronounced absorption troughs and broad emissions.
The HeI lines have become stronger so that the  $\lambda6678$  line is clearly identified in addition to the $\lambda5876$ and $\lambda7065$ lines.
In the blue strong H \& K CaII and FeII lines are visible which indicate photospheric expansion velocities \vph$=10400$ and 10000 \kms, respectively.

The next available spectrum on {\bf day +4.1, Aug. 31}, though quite noisy,  has been modeled using both SYNOW
(\tbb$=7000$ K  and \vph$=12000$ \kms, \citealp{deng}; \vph$=10000$\ kms, \citealp{branch02}) and PHOENIX \citep{ketchum08} giving an effective temperature of 6000--6500 K and \vph=10000 \kms. The features are well fit by the same ions as on day --6 (CaII, FeII, HeI and OI). The model 
with standard solar metallicity reproduces fairly well also the He lines both in absorption and in emission.
Again the best fit temperature by \citet{james10} is lower (T$_{mod}$=5250 K). 
Our black--body fit to the dereddened spectrum of day +4.1 provides  \tbb$\sim6400$ K  which is lower than the values derived for the two bracketing epochs  (\tbb$=7800$ and 7000 K for day +0.8 and 6.8, respectively) possibly because of a poor instrument response calibration.
The position of the Fe lines at about 5000 \AA\/ corresponds to \vph$=10000$ \kms, in good agreement with the velocities derived in the spectral modeling. 

Of much better quality (S/N$>100$) is the spectrum obtained on {\bf day +6.8}.
The  fit to the SED provides \tbb$=7000$K, and the expansion velocity from the FeII lines decreases to \vph$=8600$ \kms. 

In the following week the evolution slows down. Two spectra are available on {\bf day +12.5
and +14}, the former with an excellent signal--to--noise (S/N$=150$) and resolution (3.5 \AA), the latter with wider spectral range \citep{matheson}. 
Our analysis provides a similar temperature, \tbb$\sim5500$ K in both spectra, but significantly different photospheric velocities (\vph$=6400$ and 4900 \kms, respectively).
Both the Galactic NaID lines as well as those originating in the host galaxy are well resolved, as mentioned in Sect. \ref{sect:red}. HeI $\lambda 5876$ is very broad and possibly heavily contaminated by NaID.
Also the lines at $\lambda6678$ and $\lambda7065$ clearly stand out, whereas others ($\lambda4472$ and $\lambda7281$) are less pronounced (the former probably blended with MgII).  The well developed HeI lines make the spectrum closely resemble that of prototypical Type Ib supernovae.
\citet{ketchum08} have studied the spectrum of day +14: again the solar metallicity model fits the observations better, though the  CaII and HeI absorptions are too strong. TiII starts to contribute significantly to the FeII dominated region below 5000 \AA.

Three spectra have been obtained on Sept. 14 ({\bf day +17.5 to +17.8}) at different telescopes (cfr. Tab.~\ref{spec_tab} and \citealp{deng}). The SEDs of our two spectra are similar (\tbb$=4800$ K) but the S/N ratio of that obtained with EFOSC2 (plotted in Fig.~\ref{spec_fig}) is definitely higher.  The \citet{deng} spectrum, significantly bluer (\tbb$=5700$ K), has been used in the modeling both by  \citet{deng} and \citet{ketchum08}. Again the standard composition model seems to be preferred though with too strong HeI and CaII absorptions. The index $n$ of the density profile $\rho \propto r^{-n}$ was decreased from 13 to 10 to improve the fit.

The subsequent spectra ({\bf day +21 and +42}, i.e. Sept.~17 and Oct.~8) come from  \citet{matheson}. At both epochs the main features and the SEDs have not changed significantly with respect to day +18. We measure \tbb$\sim4900$ K on the dereddened spectrum, in good agreement with the temperature adopted in the modeling \citep[respectively 4800 and 4600 K,][]{branch02,james10}.
The measured expansion velocities are \vph$=5500$ and 4600 \kms, respectively. 
In the latter spectrum one can notice that the CaIR triplet starts to be resolved with an absorption at about 8360 \AA, corresponding to a velocity $v(\rm CaII)\sim7700$ \kms. 
As the temperature of the SN has decreased to about 5500K, the region bluer than 5000 \AA\/  is largely dominated by TiII lines
\citep{ketchum08}. This explains why the high metallicity model fails to reproduce the observations causing too strong absorptions.

The latest spectrum of the photospheric series ({\bf day +68}) is the average of two spectra taken with the same instrument on two consecutive nights. The HeI lines are still visible as well as the underlying continuum (\tbb$=4400$ K). Nevertheless, the progressive transition to the nebular phase can be recognized. In particular, the broad [OI] \lam\lam 6300,6364 and[CaII] \lam\lam 7291-7323 ([OII] \lam\lam 7320,7330) features start to appear clearly.

The characterizing features of the nebular spectra of SNIb/c are, indeed, [OI] and [CaII] emission lines.
These are well developed at the epoch of our last spectroscopic observation ({\bf day 371}). Once deblended into its two components, the [OI] line has an overall gaussian profile centered slightly redward of the restframe position (6309\AA) with FWHM$\sim4500$ \kms, though the relatively poor signal--to--noise ratio still allows for the presence of additional structures in the line profile \citep{tauben09}. 
In any case there is no evidence for profile distortion or a blueshift due to dust formation as detected in other objects \citep[e.g. SN~1990I,][]{elmhamdi04} in the same [OI] line.

\begin{figure}
\includegraphics[width=8.5cm]{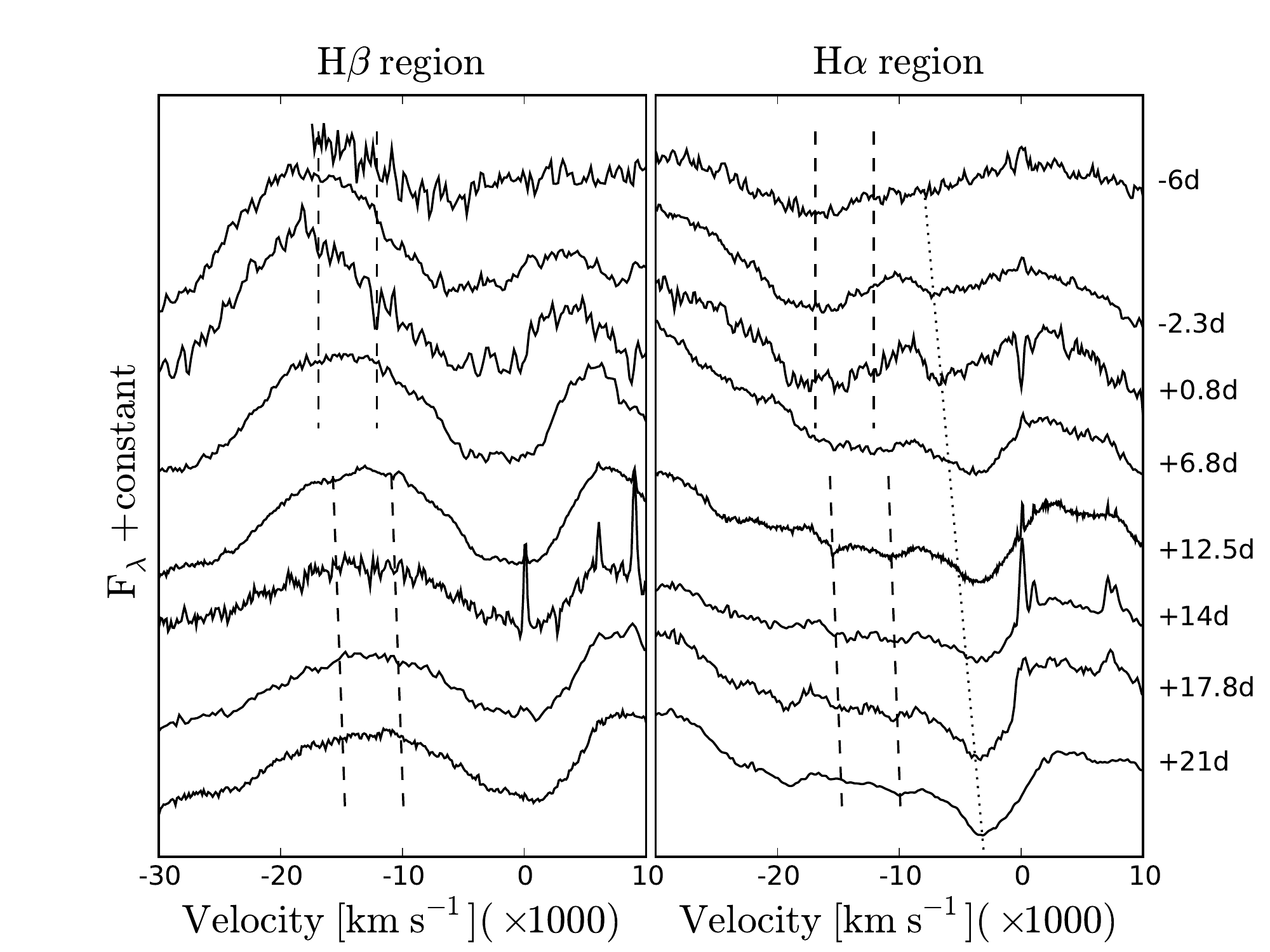}
\caption{Zoom of the 4600\AA\/  (left) and 6200\AA\/ (right panel) spectral regions during the first weeks of evolution of \dn. The x axes are in expansion velocity coordinates with respect to the rest frame positions of \Hb\/ and \Ha, respectively.  The phases relative to B maximum light are  indicated on the right axis. To guide the eye two dashed vertical lines are drawn in the spectra around maximum, corresponding to expansion velocities of about 16900 and 12100 \kms. 
After day +12.5 the broad 6200\AA\/ feature is replaced by two narrower and weaker features which slightly slow down with time. Corresponding lines are drawn also in the left panel relative to \Hb. The dotted line in the right panel is drawn at the position of HeI $\lambda6678$. }
\label{Ha_evol}
\end{figure}

\begin{figure}
\includegraphics[width=8.5cm]{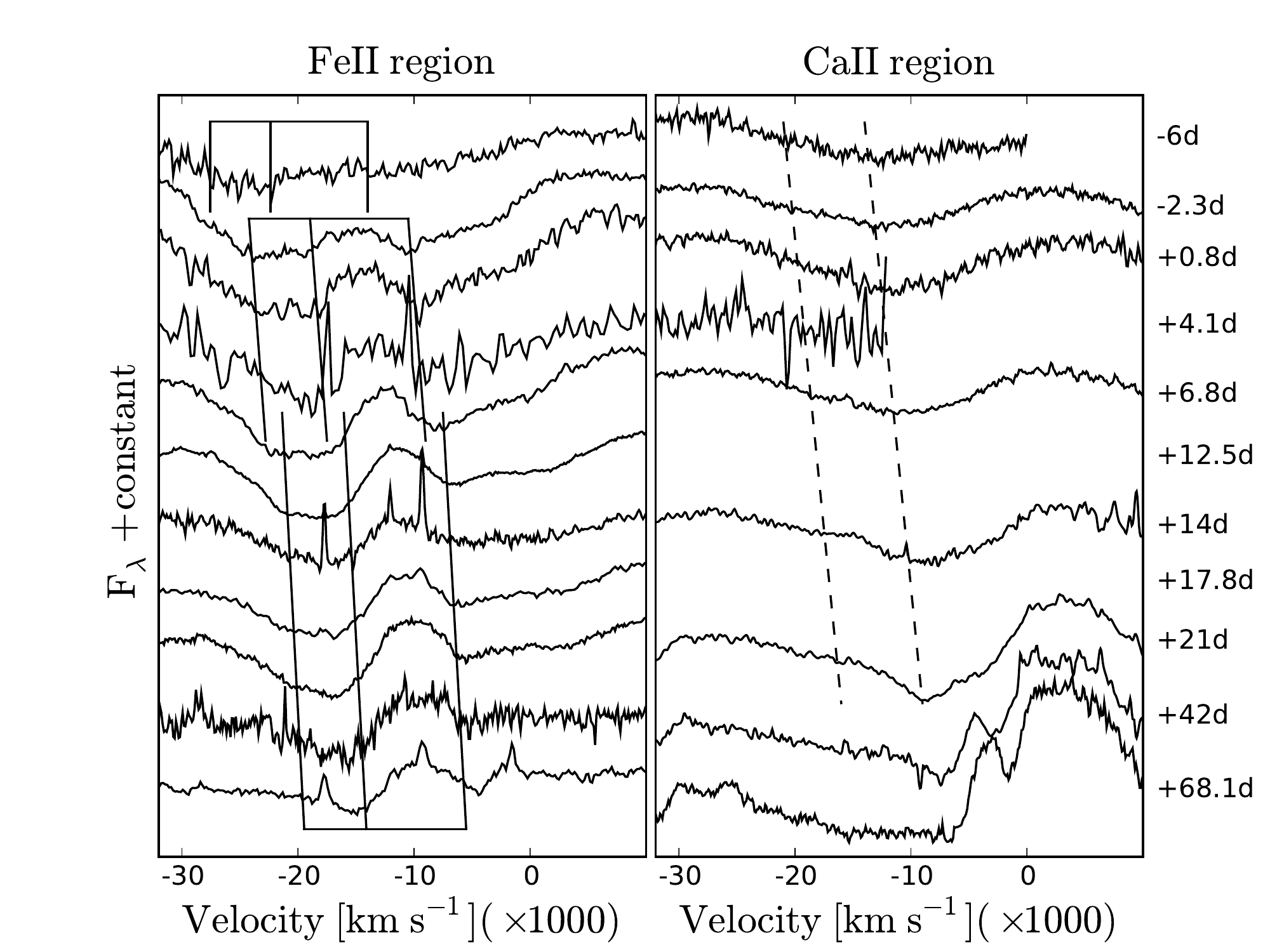}
\caption{Zoom of the FeII  (left) and CaII~IR (right panel) spectral regions during the first months of evolution of \dn.  In the left panel, the approximate positions of FeII $\lambda4924$, $\lambda5018$ and $\lambda5169$ are marked. The velocity reported on the abscissa refers to the $\lambda5169$ line and ranges from --14000 to --5500 \kms\/ from day -6 to day 68 past B maximum.
In the right panel two dashed lines are drawn to guide the eye for the main component (right) and the possible high velocity feature (left) of CaII.}
\label{CaII_IR_evol}
\end{figure}

\subsection{The expansion velocities and the presence of Hydrogen} \label{sec:Ha}

Hydrogen lines at early epochs distinguish the spectra of SNIIb from other SE SNe. Actually, faint H lines have been revealed unambiguously in the type Ib SN~2000H and, with lower confidence, in other SNIb  \citep{branch02,james10}.
These authors concluded  that H is present with low optical depths in SNIb in general and that it is located in a detached shell with velocities as high as 11000--13000 \kms. 
In this context, some attention has been paid in the previous spectral analyses of \dn\/ \citep{deng,branch02,ketchum08,james10} to a broad feature present in the spectra around maximum at about 6200 \AA.


The feature seen in the first epoch (day --6) was attributed by \citet{deng} to \Ha\/  or CII \lam 6580, arising in a detached layer with velocity v=19000 or 20000 \kms, respectively. Also \citet{branch02} interpreted the feature as due to \Ha\/ from a detached layer of  H  at 18000 \kms.
The possible alternative identification as  SiII, proposed by \citet{wooseast97} for SN~1984L, results in an expansion velocity (7300 \kms) significantly lower than that of any other ion and is, therefore, considered unlikely. 
Similar identifications were proposed for the spectra of day +4.1 (\Ha, \citealp{branch02}, possibly blended with CII, \citealp{deng}) and day +17 and +21 (CII, \citealp{deng}, or FeII, \citealp{branch02}).
The spectra were revisited by \citet{ketchum08}. They found that at early times, when the 6200 \AA\/ feature is strong and broad, an uniform atmosphere of three times solar metallicity, devoid of any H, could provide a plausible explanation, making the feature a blend of FeII lines and SiII \lam 6355. At later times the feature splits into multiple, distinct, weaker features, and solar metallicity fits better. However, they admit that higher metallicity in the outer envelope (as seen in the early time spectra) is difficult to explain, {\bf also in the light of our direct determination of a somewhat sub-solar metallicity for the SN environment (c.f.r. Sect \ref{sec:disc}).} The simpler interpretation of the 6200 \AA\/ feature, therefore, seems to be the presence of H.
The new, specific analysis by \citet{james10} confirms this identification and could suggest a H mass of M$_H\leq10^{-3}$ \M\/ in an outer shell of solar composition above the He core.

With this work we add new, high signal-to-noise spectra around maximum, and readdress the issue of the possible presence of H. 
In Fig.~\ref{Ha_evol} we have zoomed into the \Ha\/ (right panel) and \Hb\/ (left panel) regions of the best available spectra.
To guide the eye in the right panel we have drawn two vertical dashed lines corresponding to velocities of  $-16900$ and $-12100$ \kms\/ at the earliest epoch, assuming the \Ha\/ identification. 
The dotted line at lower velocity indicates the position of HeI $\lambda6678$ (the same as in Fig.~\ref{spec_fig}). The line is tilted to roughly match the velocity decrement.
Contrary to HeI, the 6200 \AA\/ feature appears at constant velocity ($-16900$ \kms\/) up to maximum light, while a notch at about $-12000$ \kms\/ might be present. Should these features be due to single lines and not to the conspiring effect of line blending (e.g. FeII and SiII), the constancy in expansion velocity is an indication of detached layers.
Note that the blue wing of the strong HeI $\lambda5876$ line extends outward to about $-18000$ \kms\/ on day --6 and $-15500$ \kms\/ on day +0.8, close to the estimated velocity of the fastest H layer.

By day +6.8 the broad 6200\AA\/ feature has disappeared.   
Starting on day +12.5 down to day +21 two distinct, resolved features spaced by about the same amount as before are clearly visible. They were noted by \citet{ketchum08} who suggested the same possible interpretation with FeII, SiII, CII and TiII. The slower component, if identified with \Ha, shows marginally higher velocity than HeI \lam 6678 and slows down with time at the same rate ($\sim1000$ \kms\/ in 10d).

To check whether this pair of lines are due to H, we have drawn two lines at corresponding velocities on the left panel of Fig.~\ref{Ha_evol}  relative to \Hb.  Although the \Hb\/ optical depth is expected to be significantly smaller than that of  \Ha, we note that possible signatures of \Hb~ are recognizable at the expected positions in the spectra of days --2.3 and  +12.5, the latter having the best signal--to--noise and spectral resolution. Though not compelling, we consider this an additional evidence that H is present in the spectrum of \dn\/ both in a detached layer above the photosphere and mixed with HeI.

A 6250\AA\/ feature was identified as \Ha\/ in the early spectra of the peculiar SN~2005bf at comparable velocity ($\sim-15000$ \kms) on the basis of the detection of similar tiny features at the corresponding position of \Hb\/ \citep{folatelli06}. Also CaII and FeII lines had  components with similar expansion velocity. For this reason we have investigated the possible presence of high velocity features for both these ions also in \dn. In Figure~\ref{CaII_IR_evol} the spectral regions of interest are plotted in analogy to those in Fig.~\ref{Ha_evol}. Broad and shallow FeII lines seem present early on also in \dn\/ at about $-14000$ \kms, a velocity exceeding that measured for HeI ($-12700$ \kms) and expected for the photosphere on day --6. 
The FeII lines are easily detected at smaller velocity in the following spectra.
Strong, broad CaII~IR triplet absorption is detected at a velocity comparable to that of  HeI. 
In the spectra of highest signal-to-noise ratio (day +6.8 and +14) one may also see another component about 5000 \kms\/ faster
but the first detection of high velocity features of CaII starting at this epoch makes the identification unlikely.
We conclude that, in addition to \Ha, there is evidence of  weak high-velocity  FeII features before maximum.

\begin{figure}
\includegraphics[width=9cm, height=10cm]{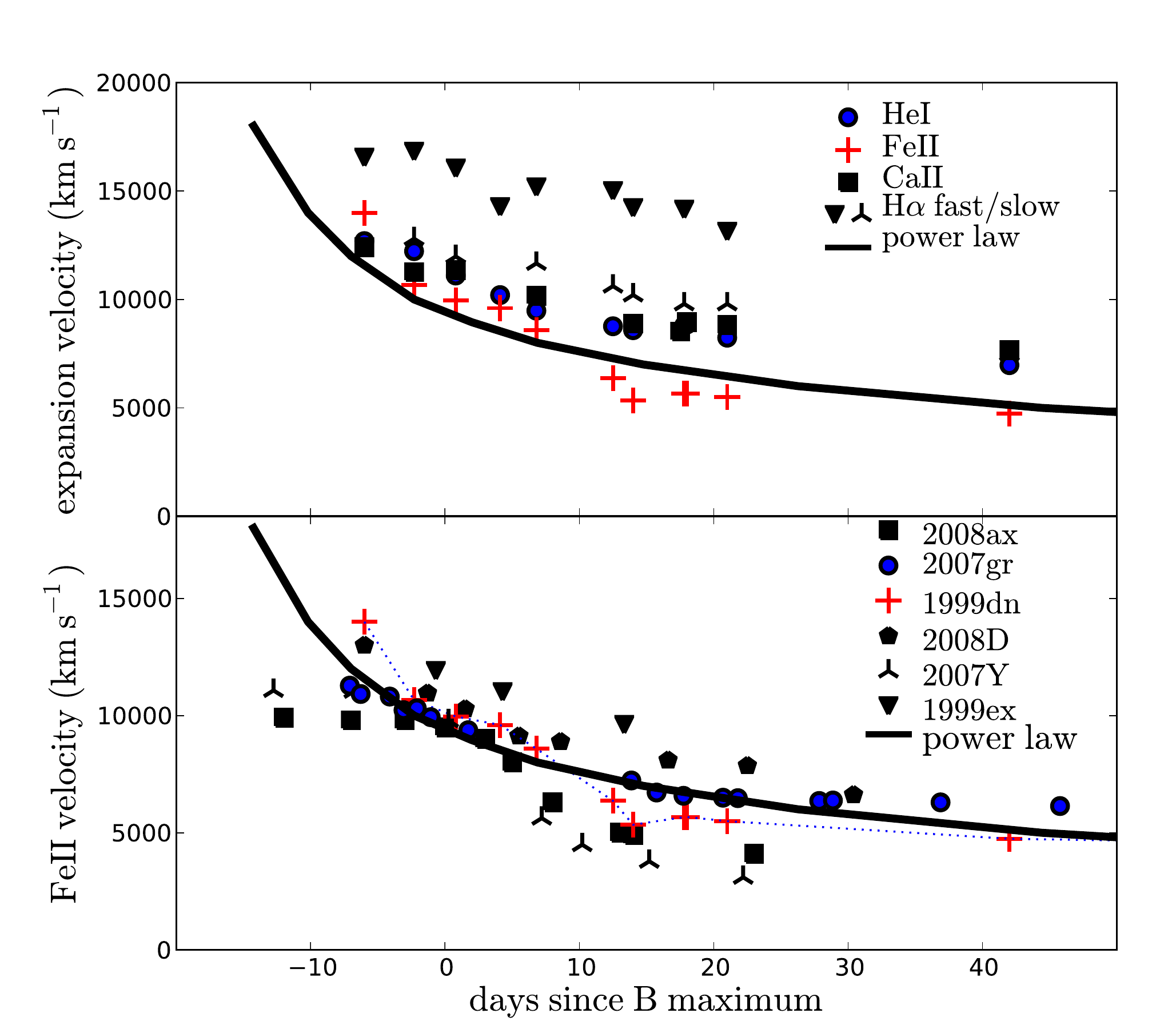}
\caption{
Top panel: velocity evolution of the HeI, FeII , CaII~IR and \Ha\/ lines of \dn\/ in the photospheric phase. 
The crosses are the averages of the velocities derived from the minima of FeII lines, adopted as indicators of the photospheric velocity; the circles the average velocities of HeI lines; the triangles (starred triangles) the velocity of the $fast$ ($slow$) component of \Ha, and the squares the velocity of the CaII IR triplet.
Bottom panel: comparison of the  velocity evolution of the FeII lines in several SE SNe.
In both panels the curve is the power-law fit from  \citet{branch02}.}
\label{vel_evol}
\end{figure}

In Fig.~\ref{vel_evol} (top panel) we summarize the evolution of the line velocities as a function of time. 
The standard deviation of the velocities derived for HeI lines is $\sigma=400-700$ \kms, possibly due to blending of various lines (e.g. NaID for \lam 5876, contamination by telluric features for \lam 7065, or by narrow \Ha\/ emission of the parent galaxy for \lam6678). 
We estimate uncertainties of the same order for the possible H lines  because of their weakness.
The photospheric velocity of \dn\/ is in agreement with those of other normal SNIb \citep{branch02} and is well fitted  by the power-law, $v_{ph}\propto t^{-2/(n-1)}$ with $n=3.6$.
As for other SNIb, HeI seems to be undetached at the first epoch but detached afterward.  The lowest velocity of the HeI layer is measured at about 6000 \kms.
The faster H is detached at all epochs with velocity ranging between $17000-13000$ \kms.  
The slower H component remains about 1500 \kms\/ faster than HeI at all epochs.
The two H components somehow bracket the H velocities of other SNIb reported in Fig.~23 of \citet{branch02}.

The bottom panel of Fig.~\ref{vel_evol} compares the FeII velocity of \dn\/ with those of other SE SNe. \dn\/ has an expansion velocity smaller than \ex\/, but higher than other objects, e.g. SN~2007Y. Only at early phases the velocity seems to deviate from the power-law fit by  \citet{branch02}. At these early phases the expansion velocity is, in fact, more similar to that shown by the energetic SN~2008D. 
The  overall normal velocity behavior  indicates that the broad (slowly evolving) light curve of \dn\/ after the peak is probably not due to a low expansion velocity but to a larger ejected mass.

\subsection{\dn\/ and other Stripped Envelope SNe}

\begin{figure}
\includegraphics[width=9cm, height=13cm]{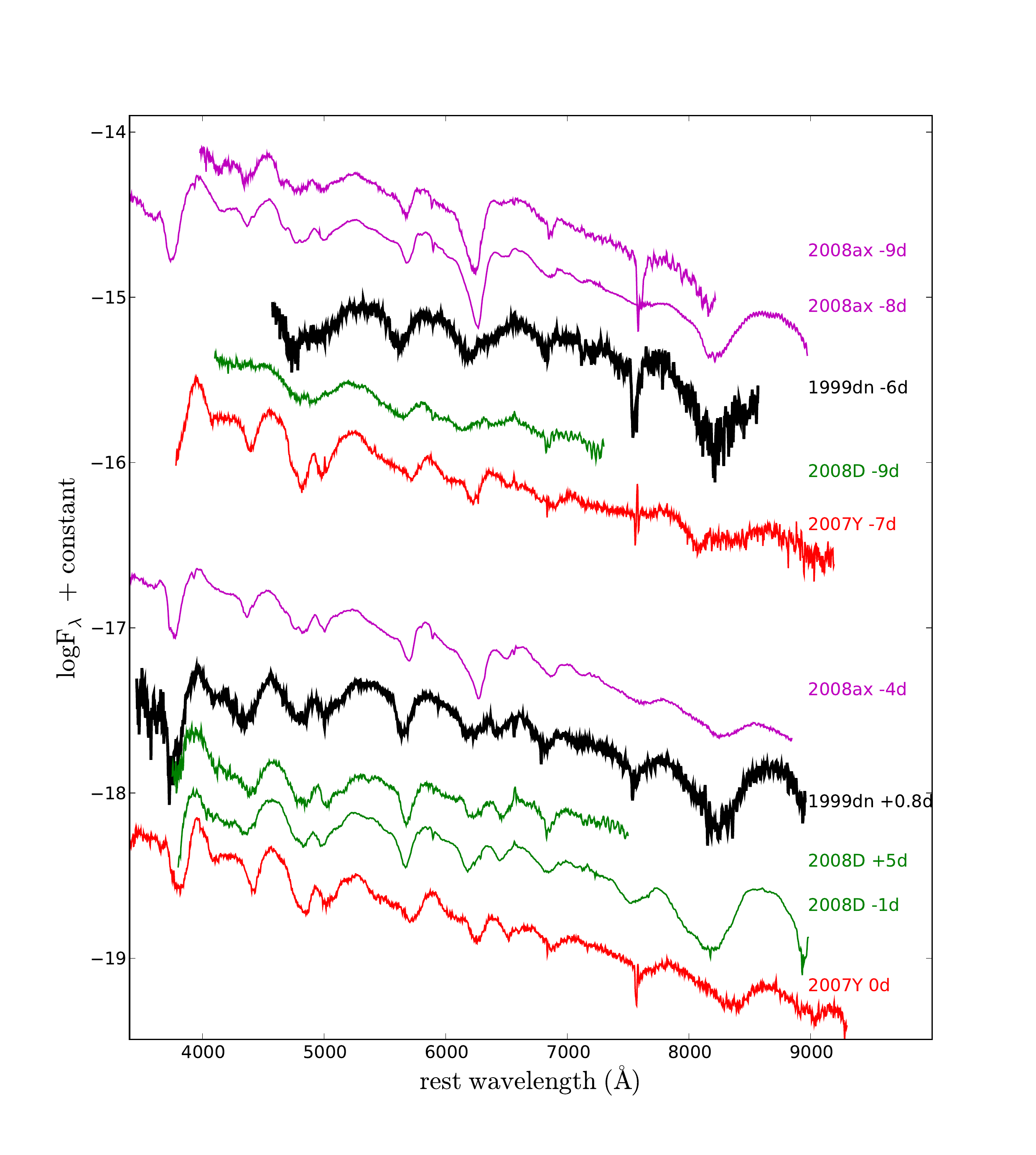}
\caption{Comparison of the pre-maximum (--1 week, top) and maximum-light (bottom) spectra of \dn\/ with those of other SE SNe. The spectra are plotted in the SN restframe and dereddened according to Tab.~\ref{comp_sample}. The phase relative to the B maximum of each object is reported close to the SN identifier. 
The spectra come from \citet{mazzali08} and \citet{valenti11} for \d, from
\citet{pasto08} and \citet{tauben10}   for \ax, from \citet{stritzinger09} for \y, and from \citet{deng} and this paper for \dn.}
\label{comp_pre}
\end{figure}

We have compared the spectra of \dn\/ with those of other SNe by means of GELATO, the spectra comparison tool developed by \citet{avik08} which compares input spectra with those present in our archive. Not unexpectedly, the best match is always with those of other SE SNe. In Fig.~\ref{comp_pre} we show the comparison with a number of well studied objects during the pre-maximum and maximum phase.
At these epochs the objects that match \dn\/ best are the extensively studied SNIIb \ax\/ \citep{pasto08,chornock10,tauben10} and the energetic SN 2008D \citep{mazzali08,modjaz09}.

For the first spectrum of \dn\/ (--6d) the closest match is with \ax\/ on days --9 and --8 \citep{pasto08,tauben10}.
The strong \Ha, which unequivocally marks the presence of H in \ax\/ at this epoch, corresponds in position to the 6200 \AA\/ feature of \dn. Note, however, that the spectra of \ax\/ that best match those of \dn\/ are at an earlier phase, probably because of a faster expansion velocity of \dn\/. At this epoch, the numerical match with \d\/ is poorer because its expansion velocity at such epoch is even higher \citep{mazzali08}.
There is also a general resemblance to \y, but the expansion velocity of \dn\/ is significantly larger.

Also at maximum \dn\/ best matches \d\/ and \ax\/ before their B maxima. The SED of \dn, however, is slightly redder.
The 6200 \AA\/ feature corresponds to the fast blue wing of the \Ha\/ absorption of \ax.

\begin{figure}
\includegraphics[width=9cm, height=13cm]{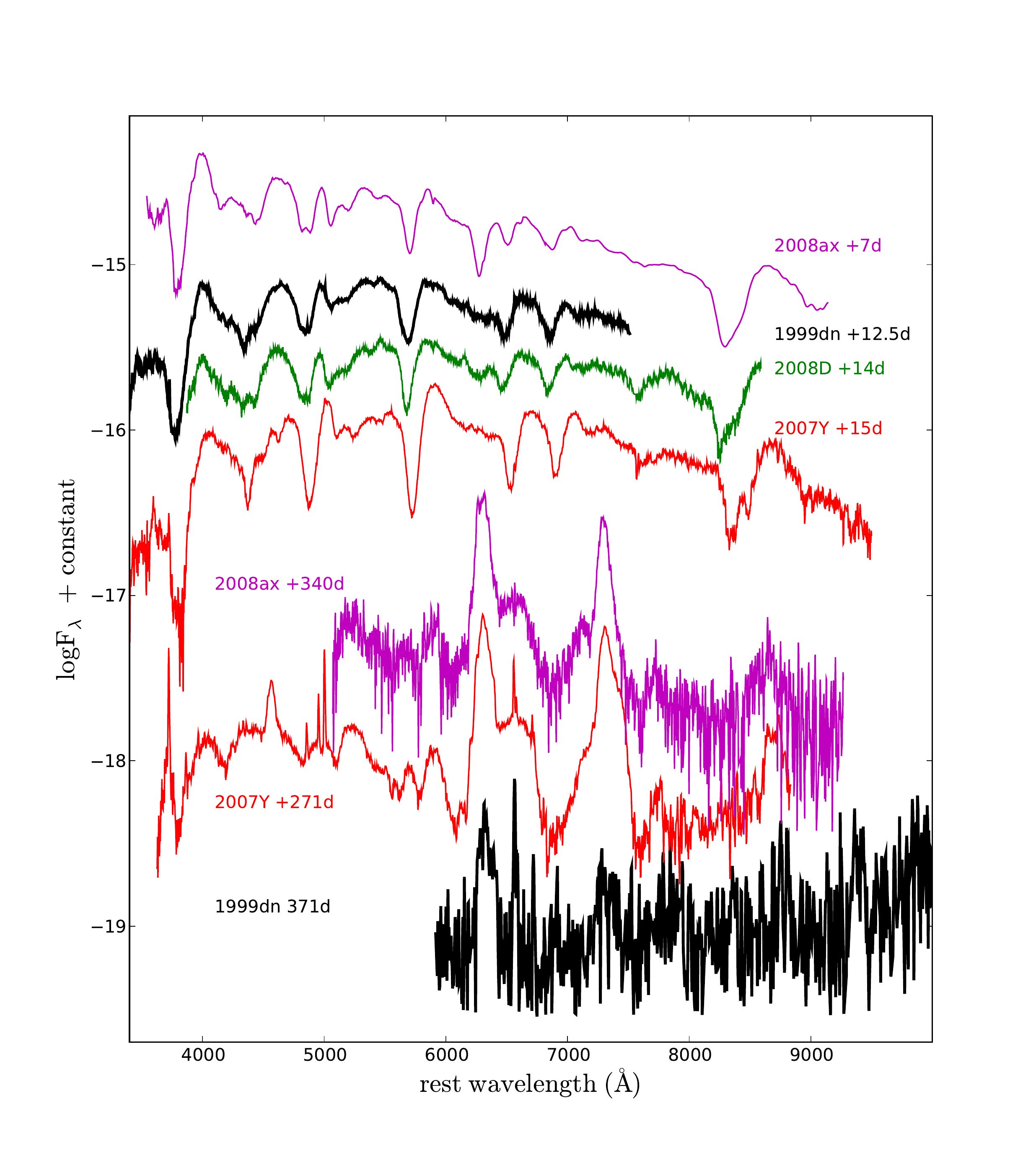}
\caption{As in Fig.~\ref{comp_pre} but about 2 weeks (top) and 1 year (bottom) post-maximum.
The spectra come from \citet{modjaz09} for \d, 
\citet{tauben10}   for \ax, \citet{stritzinger09} for \y\/ and this paper for \dn.}
\label{comp_post}
\end{figure}

The comparison after maximum is shown in Fig.~\ref{comp_post}. On day +12.5 (top)
 the spectra of all objects are very similar. 
While \Ha\/ in \ax\/ is still very strong, the 6200 features of both \dn\/ and \y\/
have dimmed and two notches are left where the broad absorption was before (cfr. Fig.~\ref{Ha_evol}). 
Again, due to smaller expansion velocity (cfr. Fig.~\ref{vel_evol}), the absorptions in \y\/ are redder than in \dn.

A late-time (371d) red-grism spectrum has been obtained for \dn\/ (Fig.~\ref{comp_post} bottom). 
In both \ax\/ and \y, a broad shoulder on the red edge of [OI] $\lambda\lambda6300$, 6364 is clearly visible, which might 
be interpreted as a sign of interaction of a fast (v$\sim10000$ \kms) shell with circumstellar material.
However, the late time interaction scenario is not fully consistent and other interpretations have been proposed, though not proved \citep[cfr.][]{tauben10}.
At the same position only unresolved \Ha, [NII] and [SII] lines from an underlying HII region are discernible in \dn, a sign that there is not (yet) CSM interaction.
[OI] $\lambda\lambda6300$, 6364 and [CaII] $\lambda\lambda7191$, 7323 ([OII] $\lambda\lambda7300$, 7330) are clearly visible, with a [CaII]/[OI] flux ratio of $0.55\pm0.10$.

\section{Discussion} \label{sec:disc}

\dn\/ was extensively studied in the last decade but our new data allow a better determination of the main parameters (cfr. Tab.~\ref{maindata}). The light curve results to be 0.5 mag brighter (\Mv$=-17.2$) than previously estimated, but still fainter than the average of SNIb/c \citep[\Mv$=-17.70$, H$_0$=72 \kms Mpc$^{-1}$, ][]{richardson06}.
The multicolor observations (Sect.~\ref{sect:phot}) have shown a relatively strong NIR flux.
This large NIR flux is intrinsic, since the EW(NaID), the broad band colors and the SED all suggest \dn\/ to be only mildly reddened (E(B-V)$=0.1$, cfr. Sect.~\ref{sect:red}). 
The NIR accounts for about 50\% of the flux during the advance photospheric phase.
The bolometric light curve reaches the same luminosity as SNe~2007gr and 1999ex, but is brigther than \y, and has a width similar to that of SN~2004aw (cfr. Fig.~\ref{fig:bolo-conf}).
It is known that the peak luminosity gives hints on the ejected  \ni\/ mass, while the width carries information on the total mass of the ejecta and kinetic energy. 
To extract in a self consistent manner information on the ejected mass (M$_{ej}$), the nickel mass, (M($^{56}$Ni)),  and the kinetic energy (E$_K$), we have employed a toy model of the bolometric light curve \citep{valenti08b}.
Our procedure is based on a two-component analytical model to account for the photospheric and nebular phases. During the photospheric phase (t$\leq 30$ d past explosion) a homologous expansion of the ejecta, spherical symmetry, and a concentration of the radioactive $^{56}$Ni exclusively in the core are assumed \citep{arnett82}. At late times (t $\geq 60$ d past explosion), the model includes the energy contribution from the $^{56}$Ni--$^{56}$Co--$^{56}$Fe  decay, following the works of \citet{sutherland84} and \citet{capp97}. 
The incomplete trapping of $\gamma$--rays and positrons has been accounted for using the \citet{clocchiatti97} prescriptions for both photospheric and nebular phases.
The bolometric light curve model suggests a $^{56}$Ni mass M($^{56}$Ni)$=0.11$ \M. A comparison with the values from Table~2 of \citet{hunter09}, obtained with similar methods, shows indeed that M($^{56}$Ni) is marginally larger than in SN~2007gr (0.076 \M), and \y\/ \citep[0.06 \M,][]{stritzinger09}, but smaller than 2004aw (0.3 \M). 
Using an average opacity of 0.06 g/cm \citep[as in ][]{valenti08b} and a scale velocity v$_{sc}$=v$_{ph}=10100$ \kms,
we obtain an ejected mass M$_{ej}=4-6$ \M\/ and kinetic energy E$_K=5.0-7.5 \times 10^{51}$ erg. 
Surprisingly, but not totally unexpected because of the broadness of bolometric light curve of SN 1999dn, these values are comparable to those obtained for the energetic SN 2008D associated to an X--ray flash \citep{mazzali08}. In fact our toy model makes use of the scale velocity of the explosion (v$_{sc}$=v$_{ph}$,  which are similar for SNe 1999dn and 2008D) without taking into account possible differences in the density profiles. 

In the case of \dn\/ this may cause an over-estimate of the physical parameters, the real values being at the lower boundary of those ranges. A more precise determination should rely on more detailed codes. A comparison between the physical parameters deduced, for a sample of well studied SNIb, with our toy model with those derived by more sophisticated codes will be given in Valenti et al 2010 (in preparation). Anyway, even considering the possibility that some physical values for SN 1999dn are slightly over-estimate, the supernova seems to belong to the group of relatively massive and energetic SNe Ib.

We noticed in Sect.~\ref{sec:Ha} that the evolution of the photospheric velocity of \dn\/ (as well as those of HeI and H) are in general very similar to those of other SNIb which implies, following \citet{branch02}, comparable density profiles, masses and kinetic energies above the photospheres. 
As noticed by those authors, the strong similarity does not leave much room for asymmetries in SNIb, which is confirmed by the [OI] line profile of \dn\/ at late time \citep{tauben09}. Only the earliest (--6d) measurement of the FeII velocity is significantly larger than the power law fit by Branch et al..

The previously available spectra of \dn\/ before and after maximum light have been extensively modeled by means of both the highly parametrized SYNOW \citep{deng,branch02} and the more sophisticated non-LTE PHOENIX codes \citep{ketchum08,james10}. The latter successfully reproduces all main features of the spectra with a homogeneous stellar atmosphere of H, He, C, N, Ne, Na, Mg, Si, Ca and Fe with the main features identified as HeI, OI, CaII and FeII at all epochs. Also the HeI absorptions were successfully reproduced including $\gamma-$ray deposited by the radioactive $\beta$ decay of \co.

Our new spectra confirm the presence of evolving features around 6200 \AA\/ during the first weeks  (Sect.~\ref{sec:spec}).
Up to maximum light the region is dominated by a broad and strong feature which, if attributed to H, implies expansion velocities of about 17000 \kms\/ (cfr. Sect.~\ref{sec:Ha}  and Fig.~\ref{Ha_evol}).
At subsequent epochs (days +6.8 to +21) our new, superior quality spectra confirm the presence of minor features \citep[cfr.][]{ketchum08} in the same region. 
Whether these components are due to H is not clear but the detection in the highest signal--to--noise spectra of weak absorptions at the same velocities with respect to \Hb\/ seems to support the existence of two H layers, one detached at velocities v $=17000$ to $13000$ \kms, comparable to that of H in other SNIb \citep{branch02}, the other only marginally faster than HeI ($\Delta(v)\sim1500$ \kms, cfr. Fig.~\ref{vel_evol}) and possibly located in the outer He layer.
The high FeII velocity at the earliest epoch (--6d) seems compatible with the existence of high--velocity H--rich layers.

\y\/ and other SNIIb show evidence of interaction of the fast expanding H-rich layer with the CSM (Fig.~\ref{comp_post}, bottom) about one year past maximum light. This is not the case in  \dn, indicating that no major mass loss episodes have occurred in the last decade before the explosion (assuming wind velocities $\sim 2000$ \kms, typical of WR progenitors).  
There is also no evidence of dust formation at any time, neither from photometric (light and color curves) nor from spectroscopic observations (line positions and shapes).

On the spectrum obtained at WHT on Sept. 9, 1999, we have measured the O3N2 index \citep{pettini04} of the region adjacent to \dn\/ along the slit of the spectrograph, corresponding to projected linear distance of $\sim 200$ pc at the adopted distance of the SN. Relation (3) of \citet{pettini04} thus provides an oxygen abundance at the SN location $12+log(O/H)=8.39\pm0.05 \pm0.25$ (where the first error is statistical and the second one is is the 95\% spread of the calibrating relation), which is in excellent agreement with the estimate of \citet{modjaz10} (8.32) and slightly lower than the solar abundance  $12+log(O/H)=8.69\pm0.05$ \citep{asplund09} and close to the average metallicity derived for the sites of a sample of fifteen SNIb, which includes also SN~1999dn, studied in Modjaz et al.

A search for the progenitor of \dn\/ has been carried out on HST archive images \citep{vandyk03}. The progenitor was not
detected down to M$_V^{\rm o} \geq -7.3$ mag and (U--V)${\rm o} \leq 2.5$ mag (these values do not change significantly with our assumptions on H$_0$ and \ebv). 
Unfortunately this determination does not strongly constraints the nature of the progenitor either as a massive, single
WR star \citep[e.g. M$_{\rm ZAMS}\geq23-25$ \M\/ at Z=0.02,][]{georgy09}, or as a star of lower initial mass in an interacting binary system \citep[e.g. M$_{\rm ZAMS}\geq14-16$ \M\/ at Z=0.02,][]{yoon10}.
However the lack of signatures of dust favors the single, massive star scenario, given the fact that while the radiation field
of single WR stars is expected to prevent dust formation in their local environment, while binarity  in WR stars seems to provide the necessary physical conditions for it \citep[][and references therein]{crowther07}.

The moderate metallicity environment ($12+log(O/H)\sim8.4$, slightly sub-solar) in which \dn\/ exploded is not inconsistent with the single scenario. In fact, the binary channel for producing WR stars in WR galaxies (as NGC~7714) is important just at lower ($12+log(O/H) < 8.2$)
metallicity \citep[e.g.][]{lopez10}, while the probability of forming single WR stars increases with the metallicity because it
is easier to reach the WR phase due to the metallicity-dependence of the
stellar wind \citep[e.g.][ and references therein]{lopez10}. The lower mass limit for having a WR progenitor decreases from $\sim40-50$ \M\/ at Z=0.004 to $\sim23-25$ \M\/ at Z=0.02 \citep{georgy09}. 
As a consequence the progenitor of \dn\/ could be a single WR star having a main sequence mass $\geq23-25$ \M.

Also the relatively small flux ratio [CaII]/[OI]$=0.55\pm0.10$, known for being constant with time \citep[cfr. the spectra collection by][]{tauben09} is consistent with a single massive star scenario. 
In fact, \citet{frans87,frans89} have shown that this ratio is a diagnostic of the core mass of the progenitor, with higher ratios indicative of smaller cores. 
The ratios measured for \J, \y\/ and \ax\/ are 0.6, 1.0 and 0.9, respectively \citep{stritzinger09,tauben10}. 
Since the core mass is strongly dependent on the progenitor ZAMS mass,
thus we have an indication that the progenitor of \dn\/ is more
massive than the above mentioned SNe.

\section{Conclusions}

We have presented detailed photometric observations and new spectra of \dn\/ from before maximum to the nebular phase.
These new data turn this object, already considered a prototypical SNIb, into one of the best observed objects of this class.

\dn\/ was a moderately faint SNIb (M$_V=-17.2$ mag) which produced 0.11 \M\/ of \ni. With a toy model 
we have estimated an ejected mass of 4--6 \M\/ with E$_{K}=5.0-7.5 \times 10^{51}$ erg. 
Due to the rough approximation of the model, these values may be slightly over-estimated. 
Our analysis on \dn\/ confirms that, contrary to early belief, a prototypical SN Ib may produce several foe of kinetic energy and eject several solar masses.

Overall the main parameters of the explosion are comparable to those of the type Ic SN~2004aw and the massive type Ib \d, 
much higher than those of the low-energy and low-ejected-mass \y. 
Higher explosion energy and ejected mass, along with the small
flux ratio [CaII]/[OI], the lack of signatures of dust formation
and the relatively high-metallicity environment  point toward a single massive progenitor (M$_{\rm ZAMS}\geq23-25$ \M). On the other hand, none of these evidences completely 
rule out the scenario of a less massive star in a binary system.

The spectra of \dn\/ at various epochs are similar to those of other SE SNe that show clear presence of H at early (type IIb SNe 1993J and 2008ax, type Ib SNe 2000H, 2007Y and 1999ex) or late (SNe 1993J, 2008ax, 2007Y) epochs. 
Such similarities, coupled to the fact that accurate spectral modeling \citep[e.g.][]{ketchum08} did not find other satisfactory explanations for the puzzling 6200\AA\/ feature, lead us to support its identification with detached \Ha. We conclude, therefore, that 
it is likely that residual H can be recognized in the spectra of most SNIb if observed sufficiently early on.

\section*{Acknowledgments}
The \dn\/ spectra published by \citet{deng} and \citet{matheson} have been 
retrived from the Suspect database (\begin{verbatim}http://bruford.nhn.ou.edu/~suspect/index1.html\end{verbatim}).
We acknowledge the usage of the HyperLeda database (\begin{verbatim}http://leda.univ-lyon1.fr\end{verbatim})\\
We thank the referee, Chornock, R., for his helpful comments. SB, MT, EC and FB are partially supported by the PRIN-INAF 2009 with the project "Supernovae Variety and Nucleosynthesis Yields".

\end{document}